\documentclass{elsart}

\usepackage{graphics}

\usepackage{amssymb}
\usepackage{amsmath}
\newcommand{\D} {\ensuremath{\vec{D}}}
\newcommand{\I} {\ensuremath{\vec{I}}}
\newcommand{\U} {\ensuremath{\vec{u}}}
\newcommand{\im} {\mathrm{i}}

\newcommand{\stress} {\ensuremath{\vec{P}}}
\newcommand{\vStress} {\ensuremath{\vec{T}}}

\renewcommand{\vec}[1] {\ensuremath{\mathbf #1}}
\newcommand{\dprod} {\,{\scriptscriptstyle \stackrel{\bullet}{{}}}\,}

\newcommand{\tr} {\ensuremath{\operatorname{tr}}}
\newcommand{\trans}[1] {\ensuremath{#1^{\operatorname{T}}}}
\newcommand{\devs}[1] {\overset{\scriptscriptstyle\circ}{#1}}
\newlength{\skewslength}
\newlength{\skewsheight}

\renewcommand{\div} {\ensuremath{\nabla\dprod}}
\newcommand{\grad}{\ensuremath{\nabla}}

\newcommand{\ddt}[1] {\ensuremath{\frac{\partial #1}{\partial t }}}

\newcommand{\M}{\vec{m}}

\renewcommand{\Re}{\mathop{\mathrm{Re}}}
\newcommand{\Ma}{\mathop{\mathrm{Ma}}}
\newcommand{\Kn}{\mathop{\mathrm{Kn}}}

\journal{arXiv}

\begin{document}
\begin{frontmatter}


\title{The structure of shock waves as a test of Brenner's
modifications to the Navier-Stokes equations}
\author{Christopher J.\ Greenshields} \and
\ead{chris.greenshields@strath.ac.uk}
\author{Jason M.\ Reese\corauthref{cor}}
\ead{jason.reese@strath.ac.uk}
\corauth[cor]{Corresponding author.}
\address{Department of Mechanical Engineering, University of Strathclyde,
\newline Glasgow G1 1XJ, UK}

\begin{abstract}
Brenner has recently proposed modifications to the Navier-Stokes equations that
are based on theoretical arguments but supported only by experiments having a fairly limited range \cite{Brenner:2005a,Brenner:2005b}.  These modifications relate to
a diffusion of fluid volume that would be significant for flows with high
density gradients. So the viscous structure of shock waves in gases should
provide an excellent test case for this new model.  In this paper we detail the
shock structure problem and propose exponents for the gas viscosity-temperature
relation based on empirical viscosity data that is independent of shock
experiments. We then simulate shocks in the range Mach 1.0--12.0 using the
Navier-Stokes equations, both with and without Brenner's modifications.  Initial
simulations showed Brenner's modifications display unphysical behaviour when the
coefficient of volume diffusion exceeds the kinematic viscosity. Our subsequent
analyses attribute this behaviour to both an instability to temporal
disturbances and a spurious phase velocity-frequency relationship.  On equating
the volume diffusivity to the kinematic viscosity, however, we find the results
with Brenner's modifications are significantly better than those of the standard
Navier-Stokes equations, and broadly similar to those from the family of
extended hydrodynamic models that includes the Burnett equations.  Brenner's
modifications add only two terms to the Navier-Stokes equations, and the
numerical implementation is much simpler than conventional extended hydrodynamic
models, particularly in respect of boundary conditions. We recommend further
investigation and testing on a number of different benchmark non-equilibrium
flow cases.
\end{abstract}

\end{frontmatter}

\section{Introduction}
\label{sec:introduction}
The generally-accepted parameter which indicates the extent to which a local
region of flowing gas is in thermodynamic equilibrium is the Knudsen number:
\begin{equation}
    \Kn = \frac{\lambda}{L} \propto \frac{\Ma}{\Re},
    \label{eq:Knudsen}
\end{equation}
where $\lambda$ is the mean free path of the gas molecules, $L$ is a
characteristic length of the flow system, the Mach number of the flow $\Ma =
|\U|/c$ with $\U$ the flow velocity and $c$ the speed of sound, and the Reynolds
number $\Re~=~\rho |\U| L / \mu$ with $\rho$ the mass density and $\mu$ the
dynamic viscosity. (For high $\Re$ flows over solid bodies, the characteristic
length scale is the boundary layer thickness, and the denominator on the right
hand side of eq.~(\ref{eq:Knudsen}) is $\sqrt{\Re}$~\cite{Schaaf&Chambre:1961}.)
As $\Kn$ increases, the departure of the gas from local thermodynamic
equilibrium increases, and the notion of the gas as a continuum-equilibrium
fluid becomes less valid.  The range of use of the continuum-equilibrium
assumption is therefore clearly limited, with the applicability of the classical
Navier-Stokes-Fourier equations (with standard no-slip boundary conditions)
confined to cases where $\Kn \lesssim 0.01$, typically. {\em Extended}, or {\em
modified}, {\em hydrodynamics} attempts to extend the range of applicability of
the continuum-equilibrium fluid model into the so-called `intermediate-$\Kn$'
(or `transition-continuum') regime where $0.01\lesssim\Kn\lesssim 1$.

Howard Brenner of MIT recently proposed modifications to the
Navier-Stokes-Fourier equations for flows with appreciable density
gradients~\cite{Brenner:2005a}.  His theoretical developments and experimental
validations are centred on slow moving flows where variations in density are
primarily caused by variations in temperature rather than pressure.  Of
particular interest to Brenner is the motion of particles due to thermal
gradients, called `thermophoresis', which provides good, yet fairly limited,
supporting evidence for his work.  This range of evidence should be broadened,
particularly since his work challenges the fundamentals of conventional fluid
dynamics and so demands rigorous validation. It is therefore of particular
interest to see whether his modifications to the classical Navier-Stokes-Fourier
equations improve their predictive capabilities for intermediate-$\Kn$ flows.

In this paper, we investigate the application of Brenner's modified
Navier-Stokes equations to the shock structure problem.  This is a
validation case used previously~%
\cite{Muckenfuss:1962,Bird:1970,Barcelo:1971,Steinhilper:1972,%
Sturtevant&Steinhilper:1972,Alsmeyer:1976,Narasimha&Das:1986,%
Lumpkin&Chapman:1991,Zhong&al:1991,Reese:1993,Levermore&Morokoff:1998,%
Cercignani&Al:1999,Myong:2001,Macrossan&Lilley:2003,Xu&Tang:2004,%
Torrilhon&Struchtrup:2004,Struchtrup:2005}
for several proposed extended hydrodynamic models, such as the Burnett
equations. Brenner's modified equations can be considered an extended
hydrodynamic model, and their relationship to established models is discussed
towards the end of this paper.  In order to be concise and avoid ambiguity, for
the remainder of this paper we shall use the expression ``Navier-Stokes'' to
refer to the classical Navier-Stokes-Fourier equation set, and adopt the term
``Brenner-Navier-Stokes'' to refer to Brenner's modified version of these.

\section{The shock structure problem}
\label{sec:shockStructure}
The shock structure problem concerns the spatial variation in fluid flow
properties across a stationary, planar, one-dimensional shock in a monatomic
gas.  We define the flow as moving at a speed $u$ in the positive $x$-direction,
with the shock located at $x = 0$; the upstream conditions at $x = -\infty$ are
super/hypersonic and denoted by a subscript `1', downstream conditions at $x =
+\infty$ are denoted by a subscript `2'.  While shocks are often modelled as
discontinuities, their physical properties in fact vary continuously from their
upstream to their downstream levels over a
characteristic distance of a few mean free paths because the relaxation times
for heat and momentum transport are finite.  The flow in this shock layer is far
from being in local thermodynamic equilibrium, typically $\Kn = 0.2 \sim 0.3$
i.e.\ very much within the intermediate-$\Kn$ regime.

Since the Navier-Stokes equations perform poorly at these $\Kn$, the shock
structure problem is particularly apposite for testing extended hydrodynamic
models.  The problem possesses certain features that make it attractive for
numerical investigation, particularly if hydrodynamic models with high-order
derivatives are used:
\begin{enumerate}
\item there are no solid boundaries to consider;
\item the upstream and downstream boundary conditions are clearly defined
through the Rankine-Hugoniot relations, with all gradients of flow quantities
tending to zero far upstream and downstream of the shock;
\item the problem is one-dimensional and steady.
\end{enumerate}
A monatomic gas possesses no modes of vibration or rotation, cannot dissociate,
and only ionizes at the highest temperatures, so its thermodynamic behaviour is
generally much simpler than that of polyatomic gases.  It is for this reason
that monatomic gases have generally been preferred as the test gas in shock
structure experiments and analysis.  Resulting data have historically been
presented in normalised form and the lack of access to the raw data presents an
opportunity for us to specify the test problem in a nondimensionalised form that
reduces the pre- and post-processing effort\footnote{In what follows, the use of
a power law viscosity model, described in section \ref{sec:powerLawViscosity},
is anticipated.}:
\begin{itemize}
\item It is convenient to set the upstream temperature $T_{1}=1$ because, for a
given $\mu_{1}$, the coefficient of proportionality $A$ is then independent of
the exponent of $s$ in the power law relation of
eq.~(\ref{eq:viscosityPowerLaw}). For simplicity we then choose $A=1$ and,
hence, $\mu_{1}=1$.
\item It is useful to set $c_{1}=1$ so that the upstream flow speed $u_{1}$
relates directly to the upstream Mach number $\Ma_{1}$.  The ratio of specific
heats, $\gamma$, is $5/3$ for monatomic gases, so the gas constant $R =
c^{2}/(\gamma T) = 3/5$ in our adopted nondimensionalised set of units.
\item The final parameters can be set so that a unit length corresponds to the
upstream Maxwellian mean free path:
\begin{equation}
\lambda_{M1}= \frac{16}{5\sqrt{2\pi\gamma}}\frac{\mu_{1}c_{1}}{p_{1}}.
\label{eq:meanFreePathMaxwell}
\end{equation}
where $p_{1}$ is the upstream pressure.  For this test case
$16/5\sqrt{2\pi\gamma} = 0.99 \approx 1$, so $\lambda_{M1} \approx
\mu_{1}c_{1}/p_{1}$.  By setting $p_{1}=1$, the unit length then corresponds
almost exactly to $\lambda_{M1}$.
\end{itemize}
All normal gradients of $p$ and $T$ are specified as zero at the solution domain
boundary far downstream.  The shock is maintained stationary and fixed within
the domain by application of the Rankine-Hugoniot velocity relation at the
downstream boundary, which for our case is
\begin{equation}
\frac{u_{2}}{u_{1}} = \frac{1}{4} \left( 1 + 3{\Ma}_{1}^{-2} \right).
\label{eq:RankineHugoniotVelocityTestCase}
\end{equation}

The case is initialised with a step change in fields from upstream to downstream
at $x = 0$.  To minimise the time required to reach a steady-state solution, the
downstream pressure and temperature are initialised using Rankine-Hug\-on\-iot
relations for pressure and temperature, which for our case are:
\begin{equation}
\frac{p_{2}}{p_{1}} = \frac{1}{4} \left( 5{\Ma}_{1}^{2} - 1 \right)
\hspace{2em} \mathrm{and} \hspace{2em}
\frac{T_{2}}{T_{1}} = \frac{p_{2}}{p_{1}} \frac{u_{2}}{u_{1}} .
\label{eq:RankineHugoniotTemperaturePressure}
\end{equation}

The initial and boundary conditions for the actual solution variables, described
in section~\ref{sec:modifiedNavierStokes} below, are simply derived from those
for $p$, $T$ and $\U$, e.g. initial and boundary conditions for $\rho$ are
calculated from $p$ and $T$ using the perfect gas law, $p = \rho R T$.  When we
include the second derivative of $\rho$ through application of
eqn.~(\ref{eq:NewtonianFluidVolumeVelocity}), below, we specify additionally
that the normal gradient of the gradient of $\rho$ is zero.

\section{The viscosity-temperature relation}
\label{sec:viscosityTemperature}
One of the main uncertainties in the physical modelling of the shock structure
problem is the relation between $\mu$ and $T$.  This is unfortunate because the
$\mu(T)$ relation has an appreciable effect on the profile of a simulated shock
--- in the extreme case of assuming a constant $\mu$, results of simulations are
very poor.  It could even be argued that the Navier-Stokes equations could be
made to work for the shock structure problem simply by adjusting the $\mu(T)$
relation until experimental shock profiles are reproduced.  The value of the
shock structure problem as a good test for hydrodynamic models therefore relies
on establishing a good $\mu(T)$ relation from reliable experimental sources,
preferably independent of shock data.  We therefore review sources to establish
$\mu(T)$ relations for a range of monatomic gases, from which argon is chosen
for our test problem.

\subsection{The power law relation}
\label{sec:powerLawViscosity}
The viscosity of a perfect rarefied gas is defined through
\begin{equation}
\mu = \tau_{1}p \propto \tau_{1}n T = \frac{5}{4}\tau_{0}n T ,
\label{eq:viscosityRarefiedGas}
\end{equation}
where $n$ is the gas molecular number density, $\tau_{1}$ is the collision
interval for momentum transport, and $\tau_{0}$ is the collision interval for
hard-sphere molecules. That the viscosity given in
eq.~(\ref{eq:viscosityRarefiedGas}) is, in fact, purely dependent on temperature
and not mass density, arises from the nature of the intermolecular force law
which determines how molecules interact in collision with each other. For
reasons of simplicity, this force is often modelled, for a given species of
molecule at a particular temperature, as varying with distance from the
molecular centre as an inverse power law with coefficient $\nu$. In a collision,
molecules approach each other with a relative speed $g$ and slow to a stop a
distance $d$ from each other when their kinetic energy is transformed to
potential energy in the force field, i.e.\ $g^{2} \propto d^{-\nu + 1}$.  With
translational temperature a function of the square of the molecular velocity, it
is then clear that the effective molecular diameter, $d \propto T^{-1/(\nu -
1)}$.  The collisional relaxation time $\tau_{0}$ is then the mean free path
($\propto n^{-1}d^{-2}$), divided by the mean molecular velocity ($\propto
T^{1/2}$) so that
\begin{equation}
n\tau_{0} \propto T^{-1/2}d^{-2} = T^{s-1} ,
\hspace{2em} \mathrm{where} \hspace{2em}
s = \frac{1}{2} + \frac{2}{\nu - 1} .
\label{eq:viscosityRarefiedGasDeriv}
\end{equation}
Equations~(\ref{eq:viscosityRarefiedGas}) and
(\ref{eq:viscosityRarefiedGasDeriv}) yield the well-known relation
\begin{equation}
\mu \propto T^{s}
\hspace{2em} \mathrm{or} \hspace{2em}
\mu = A T^{s} ,
\label{eq:viscosityPowerLaw}
\end{equation}
where $A$ is a constant of proportionality. There are two theoretical limiting
cases for the intermolecular force law: $\nu = \infty$, $s = 1/2$ corresponds to
hard-sphere molecules; $\nu = 5$, $s = 1$ corresponds to so-called Maxwellian
molecules. Real molecules generally have a value of $\nu$ ranging from about 5
to around 15.

\subsection{Experimental data for monatomic gases}
\label{sec:experimentalDataViscosity}
Increasing the value of the exponent $s$ in eq.\ (\ref{eq:viscosityPowerLaw}) in
a shock structure calculation introduces more dissipation, particularly at the
high-temperature, downstream side of the shock.  This additional dissipation
acts to smooth out the shock layer, increasing its thickness and, in particular,
lengthening the downstream ``tail'' in the flow property profiles.

Therefore, in order to test any hydrodynamic model it is important to set the
exponent $s$ independently of the shock structure problem under investigation.
In 1972, Maitland and Smith \cite{Maitland&Smith:1972} critically assessed the
viscosities of a number of gases, obtained from several different sources using
a variety of techniques, such as the capillary flow method, oscillating disc and
rotating cylinder methods, and observations of the retardation of an oil drop in
free-fall through a gas. For the monatomic gases argon, helium and xenon they
produced viscosity data which they estimated to be accurate to 1.5\% in the
temperature range 80--2000K. That upper limit of 2000K is the downstream
temperature of a shock of Mach 4.3 propagating into a room-temperature gas;
hence these viscosity data are applicable to shocks of $\Ma \le 4.3$.

In 1958, Amdur and Mason \cite{Amdur&Mason:1958} estimated the intermolecular
potential at higher temperatures from observations of the scattering of
high-velocity molecular beams, and produced tables of the viscosity of gases at
temperatures up to 15000K, corresponding to a shock of Mach 12.5. It is not
known how accurate these data are; Amdur and Mason estimated that at the higher
temperatures the error in viscosity could be as much as 10\%.

\begin{figure}[ht]
    \centering
    \input{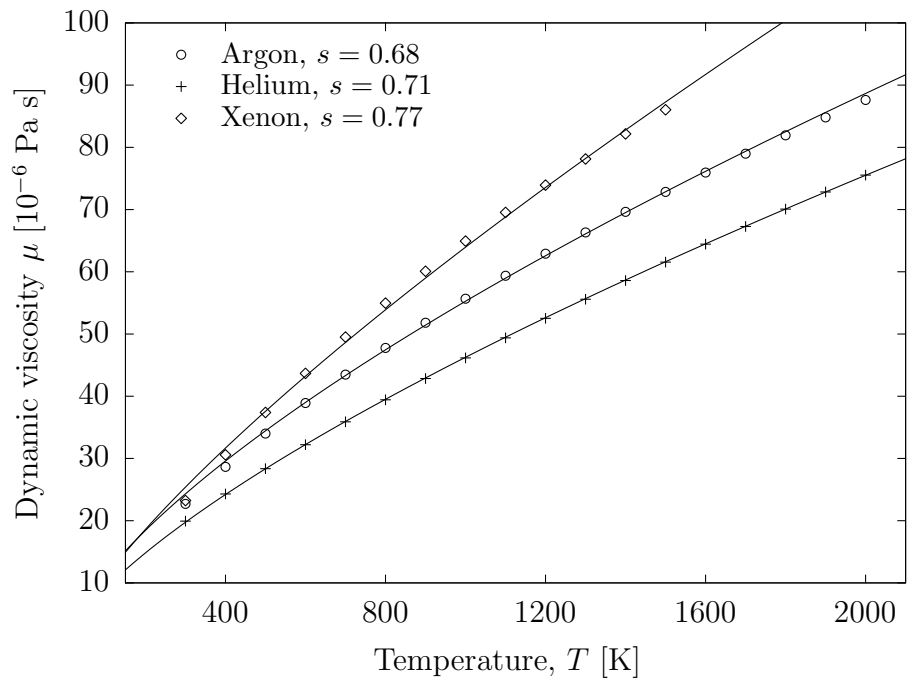}
    \caption{Experimental viscosity versus temperature data for intermediate
    temperatures for argon, helium and xenon,
    fitted to a power-law: $\mu \propto T^{s}$.}
    \label{fig:MonatomicGasViscosityIntermediate}
\end{figure}

\begin{figure}[ht]
    \centering
    \input{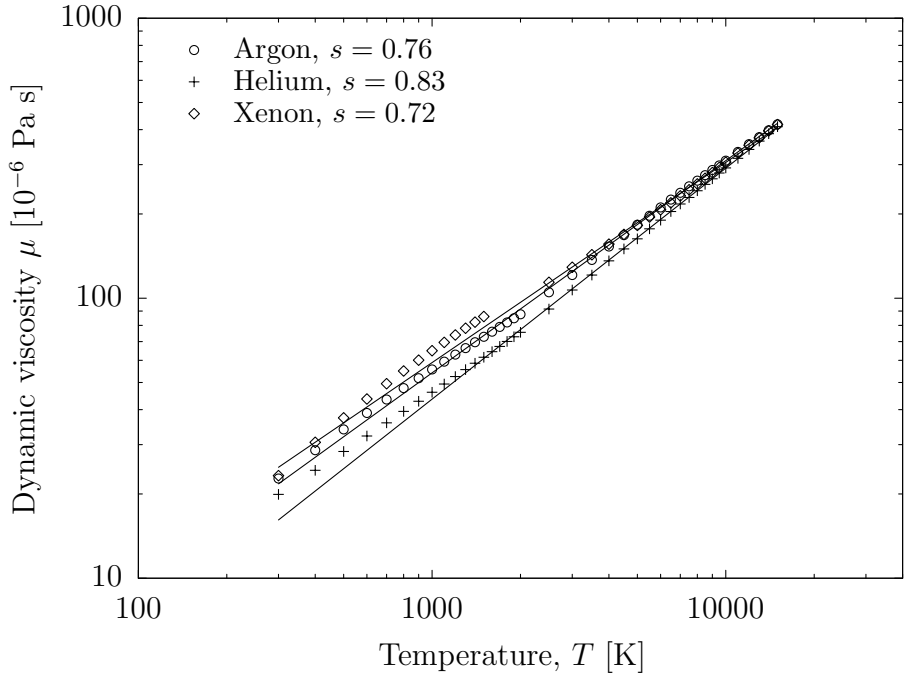}
    \caption{Experimental viscosity versus temperature data for high
    temperatures for argon, helium and xenon,
    fitted to a power-law: $\mu \propto T^{s}$.}
    \label{fig:MonatomicGasViscosityHigh}
\end{figure}

We have used these data to estimate the value of $s$ (and hence $\nu$) in eq.\
(\ref{eq:viscosityRarefiedGasDeriv}) for different temperature ranges. We have
fitted curves of the power law in eq.~(\ref{eq:viscosityPowerLaw}) by minimising
the error in viscosity for the two sets of experimental data. More details of
this process can be found in \cite{Reese:1993}, but the experimental data and
best-fit curves for three common monatomic gases are shown in
figs.~\ref{fig:MonatomicGasViscosityIntermediate} and
\ref{fig:MonatomicGasViscosityHigh}, and the corresponding exponents $s$ and
$\nu$ are given in Table~\ref{tab:experimentalViscosityTempExponents}.

The coefficient $\nu$ is itself a function of temperature because at higher
temperatures molecules with more energy can penetrate each others' force-fields
more effectively. Therefore, due to the differences in temperature range and
experimental techniques reported in the Maitland \& Smith and Amdur \& Mason
papers, we have considered two ranges of temperature or, equivalently, shock
Mach number: up to Mach 4.3, and up to Mach 12.5. (It should be noted that
the power-law fit is not very good for the high temperature data, which is
itself of unknown accuracy, therefore the $\mu(T)$ obtained must be treated with
caution.)

In the more limited temperature range of 3500--8500K, Aeschliman and Cambel
\cite{Aeschliman&Cambel:1970} obtained values for argon viscosity to an accuracy
of 12\% which can be represented to within 1\% error by a viscosity-temperature
exponent of $s = 0.74$. This value compares well with our value of $s = 0.76$ in
Table~\ref{tab:experimentalViscosityTempExponents} for temperatures in the range
2000--15000K.

Correlations between direct simulation Monte-Carlo (DSMC) simulations and
experimental shock density profiles can also provide data for the exponent $s$,
particularly at high temperatures~%
\cite{Bird:1970,Barcelo:1971,Steinhilper:1972,Sturtevant&Steinhilper:1972,%
Alsmeyer:1976,Lumpkin&Chapman:1991,Macrossan&Lilley:2003}.
While it is clearly preferable in our study of the shock structure problem to
use data that are independent of the problem itself, it is worth noting that
each of these published papers produces a value of $\nu$ that falls within a
range $9 \le \nu \le 11$, corresponding to $0.70 \le s \le 0.75$. The value $s =
0.72$ of Alsmeyer \cite{Alsmeyer:1976} is often quoted, e.g.\ recent results of
simulations using this value of $s$ agreed with experiment within an estimated
uncertainty of 5\% for temperatures above 2000K \cite{Macrossan&Lilley:2003}.

In our present study, three values of the exponent $s$ for argon are therefore
used:
\begin{itemize}
\item $s=0.68$, which is our best-fit for shock Mach numbers up to $4.3$;
\item $s=0.76$, our best-fit up to Mach $12.5$; and
\item $s=0.72$, which is the mean of our best-fit values, as well as the
commonly used value of Alsmeyer \cite{Alsmeyer:1976} that falls in the middle of
the range of values from DSMC correlations with shock density profiles.  As this
is the mean value, it is used as the `control' in our study.
\end{itemize}

We focus on the power-law form of $\mu(T)$ in this paper in order to discern
effects on shock structure due to different constitutive models for momentum and
energy diffusion rather than due to the presumed relationship between gas
properties.  However, we recognise that there are other models available for
$\mu(T)$, e.g.\ Sutherland's Law, that are generally equivalent to adding a weak
attractive component to the intermolecular force --- which is physically more
realistic.  In our simple power-law model, this attractive force would manifest
itself as an exponent $s$ that decreases as temperature increases, which is
generally reported (see, e.g., \cite{Chapman&Cowling:1970}).  It is interesting
to note, however, that our present analysis of experimental data does not appear
to bear this out: in the case of xenon in
Table~\ref{tab:experimentalViscosityTempExponents}, $s$ decreases with
increasing temperature, but with argon and helium the exponent increases with
temperature.

\section{Brenner's modification to the Navier-Stokes equations}
\label{sec:modifiedNavierStokes}

Brenner presents his modification to the Navier-Stokes equations as a change in
Newton's viscosity law.  Before arriving at that discussion, we first express
the standard governing equations in an Eulerian frame of reference as

{\bf Conservation of mass}
\begin{equation}
\ddt{\rho} + \div(\U_{m}\rho) = 0 , \label{eq:mass}
\end{equation}
{\bf Conservation of momentum}
\begin{equation}
\ddt{\M} + \div(\U_{m}\M) + \div \stress = 0 , \label{eq:momentum}
\end{equation}
{\bf Conservation of total energy}
\begin{equation}
\ddt{E} + \div(\U_{m} E) + \div\vec{j}_{e} + \div(\stress\dprod\U_{m}) = 0 ,
\label{eq:totalEnergy}
\end{equation}
where $\U_{m}$ is the mass velocity, the momentum density $\M = \rho\U_{m}$, the
total energy density $E = \rho (e + |\U_{m}|^{2}/2)$ with $e$ the specific
internal energy, $\vec{j}_{e}$ is the diffusive flux density of internal energy,
and $\stress$ is the diffusive flux density of momentum --- the familiar stress
tensor --- defined here as positive in compression: $\stress = p\I + \vStress$,
where $\vStress$ is the viscous stress tensor and $\I$ the unit tensor.

Based on this sign convention, the constitutive model for a Newtonian fluid
relates $\vStress$ to the rate of deformation tensor, $\D$, by $\mu$ and the
bulk viscosity $\kappa$:
\begin{equation}
\vStress = - 2\mu\devs{\D} - \kappa\tr(\D)\I ,
\label{eq:NewtonianFluid}
\end{equation}
where $\D \equiv \overline{\grad \U} \equiv \frac{1}{2} \left[ \grad \U +
\trans{(\grad \U)} \right]$ for a velocity $\U$, i.e.\ the overbar indicates the
symmetric part of a tensor. The deviatoric component of the deformation is
$\devs{\D}\equiv \D - \frac{1}{3} \tr (\D) \I$.

In this constitutive model, eq.\ (\ref{eq:NewtonianFluid}), it is generally
considered that $\U$ is the mass velocity $\U_{m}$ that, in the mass continuity
equation, relates to a convective flux of mass $\mathrm{d}\vec{S} \dprod
\rho\U_{m}$ at an element of surface area $\mathrm{d}\vec{S}$, or that in the
Boltzmann equation represents the statistical mean value velocity.  However,
this assumption has recently been questioned by Brenner
\cite{Brenner:2005a,Brenner:2005b} who postulated that the velocity appearing in
Newton's viscosity law should instead be the volume velocity $\U_{v}$, so named
since it relates to the flux of volume rather than mass.

The distinction between mass and volume flux is perhaps best explained by
considering a single species fluid at a molecular level.  The mass flux through
$\mathrm{d}\vec{S}$ is the product of the molecular mass and the number of
molecules passing through $\mathrm{d}\vec{S}$ in one second.  There is no net
mass flux due to random motions of molecules; at a continuum level, there is no
diffusive flux of mass, only the convective flux already defined.

To understand volume flux we can consider attributing to each molecule locally a
microscopic
portion of the volume of fluid.  The molecular volume is transported with the
molecule but will change depending on the mass density of its surroundings,
e.g.\ the microscopic
volume shrinks as the molecule moves into a denser region.
A convective flux of volume is associated with bulk motion of the molecular
volumes, and is equivalent to the ratio of convective mass flux to mass density,
i.e.\ $\mathrm{d}\vec{S}\dprod \U_{m}$.
If the fluid density varies across
$\mathrm{d}\vec{S}$, as a molecule passes through $\mathrm{d}\vec{S}$ there is
a change in its
associated volume, thus a net flux of volume.  Random motions can therefore
produce a net flux of volume,
so that there exists a \emph{diffusive} flux of volume in regions of
non-zero density gradient.
The volume flux $\mathrm{d}\vec{S} \dprod \U_{v}$ therefore represents the total
flux of volume, comprising the convective flux $\mathrm{d}\vec{S}\dprod \U_{m}$
{\em and} a diffusive flux $\mathrm{d}\vec{S} \dprod \vec{j}_{v}$, where
$\vec{j}_{v}$ is the diffusive volume flux density, such that
\begin{equation}
\U_{v} = \U_{m} + \vec{j}_{v} .
\label{eq:velocityRelation}
\end{equation}
For a single component fluid undergoing heat transfer, Brenner proposed a
constitutive equation for $\vec{j}_{v}$:
\begin{equation}
\vec{j}_{v} = \alpha_{v} \frac{1}{\rho}\grad\rho,
\label{eq:jvConstitutive}
\end{equation}
where $\alpha_{v}$ is termed the `volume diffusivity' \cite{Brenner:2005a}.
Exactly how $\alpha_{v}$ should be quantified for a given fluid state is an open
question. Brenner relates $\alpha_{v}$ directly to well-known diffusivity
coefficients under some limited conditions; in particular, for single component
fluids undergoing heat transfer, $\alpha_{v}$ is the same as the thermal
diffusivity $\alpha = k/\rho c_{p}$, where $k$ is the thermal conductivity and
$c_{p}$ is the specific heat capacity at constant pressure.

While the distinction between volume and mass velocities has been made, the
question remains of why the velocity appearing in Newton's viscosity law should
be the volume velocity rather than mass velocity.  Brenner's justification is
based on limited evidence (e.g.\ comparison of analytical solutions with
thermophoresis experiments). A lack of theoretical physical argument could
therefore
lay the hypothesis open to some criticism.  However, some support for it can be
found within the phenomenological GENERIC theory presented by
\"{O}ttinger~\cite{Oettinger:2005}.
First,
he demonstrates that the GENERIC formulation arrives at the standard
Navier-Stokes equations when the terms associated with mass density in the
friction matrix are identically zero. Then, by including non-zero terms
associated with mass density in the friction matrix, a revised set of governing
equations is derived that includes two velocities, similar to $\U_{m}$ and
$\U_{v}$ defined through eqs.~(\ref{eq:velocityRelation}) and
(\ref{eq:jvConstitutive}).  What emerges is that the standard governing
equations have historically ignored mass diffusivity on the basis that the
diffusive mass flux is zero, while forgetting that there are associated momentum
and energy fluxes that may not be zero. GENERIC includes these momentum and
energy fluxes, both of which are entropy producing, making the process of mass
diffusion irreversible.  The ability of mass diffusion to produce entropy is,
according to \"{O}ttinger, something that is missing from the conventional
Navier-Stokes equations.

Brenner's modification, essentially eq.\ (\ref{eq:velocityRelation}), can be
incorporated into the system of governing fluid equations either by recasting
the equations using $\U_{v}$ as the convective velocity instead of $\U_{m}$, or
by using $\U_{v}$ in the constitutive equation for Newton's viscosity law. The
former approach has been adopted elsewhere
\cite{Oettinger:2005,Bardow&Oettinger:2006} but here we choose the latter simply
so that the Brenner modification appears more clearly as a new extended
hydrodynamic model, rather than a radical change to the governing equations
themselves.

For a monatomic gas, $\kappa = 0$.  Combining eqs.~(\ref{eq:NewtonianFluid}),
(\ref{eq:velocityRelation}) and (\ref{eq:jvConstitutive}) yields a modified
expression for the viscous stress:
\begin{equation}
\vStress = - \mu \left[ \grad\U_{m} + \trans{(\grad\U_{m})} -
\frac{2}{3}\div\U_{m} \right]
- 2\mu\devs{ \overline{ \grad \left\{ \frac{\alpha}{\rho}\grad\rho \right\} }} .
\label{eq:NewtonianFluidVolumeVelocity}
\end{equation}
The Brenner approach requires the transport of energy to be similarly modified
through consideration of the diffusion of internal energy.  It is usually
assumed that the diffusion of internal energy consists solely of heat diffusion,
so that the diffusive internal energy flux density, $\vec{j}_{u}$, is considered
synonymous with diffusive heat transfer, $\vec{q} = -k\grad T$, according to
Fourier's law. However, the presence of a diffusive volume flux, of flux density
$\vec{j}_{v}$, enables energy to be transported across a surface by diffusive
work transfer of an amount $-p\vec{j}_{v}$. The diffusive internal energy flux
density is therefore given by:
\begin{equation}
\vec{j}_{e} = \vec{q} - p\vec{j}_{v} ,
\label{eq:internalEnergyFluxDensity}
\end{equation}
which, following our argument above, can be re-written in the form:
\begin{equation}
\vec{j}_{e} = - k\grad T - \alpha_{v}\frac{p}{\rho}\grad\rho.
\label{eq:internalEnergyFluxDensityFinal}
\end{equation}

Equations~(\ref{eq:NewtonianFluidVolumeVelocity}) with
(\ref{eq:internalEnergyFluxDensityFinal}) comprise Brenner's modifications to
the classical Navier-Stokes-Fourier equations
\cite{Brenner:2005a,Brenner:2005b}. We should note that this new fluid model has
yet to receive either independent theoretical justification or experimental
confirmation. As with any new hypothesis or model it is also subject to
refinement and re-casting into different forms.  However, we use it in this
paper in the form presented in \cite{Brenner:2005a,Brenner:2005b}, without
prejudice, to provide an indication of its current utility and limitations.

\section{Numerical solution of the governing equations}
\label{sec:solutionMethod}
Our numerical shock structure solver is developed using the open source Field
Operation and Manipulation (OpenFOAM) software \cite{openfoam}. Written in C++,
OpenFOAM uses finite volume (FV) numerics to solve systems of partial
differential equations ascribed on any 3-dimensional unstructured mesh of
polygonal cells.  All solvers developed within OpenFOAM are therefore
3-dimensional, but can be used for 1- or 2-dimensional problems by the
application of particular conditions on boundaries lying in the plane of the
direction(s) of no interest.

Fluid flow solvers in OpenFOAM are generally developed within an implicit,
pressure-velocity, iterative solution framework.  The solver we developed for
this work first solves eqs.\ (\ref{eq:mass}), (\ref{eq:momentum}) and
(\ref{eq:totalEnergy}) for $\rho$, $\M$ and $E$ respectively. The equations are
treated in a segregated manner: for each equation, terms including the solution
variable are, wherever possible, treated implicitly, with other terms treated
explicitly.  All equations include convection of transported variables that
require a consistent, conservative set of fluxes of $\U_{m}$.  After solving the
sequence of segregated equations for $\rho$, $\M$ and $E$, an iterative
PISO-style method \cite{Issa:1986} solves an equation for pressure $p$, derived
from the perfect gas law, and eqs.\ (\ref{eq:mass}) and (\ref{eq:momentum}), to
produce conservative fluxes of $\M$ from which the fluxes of $\U_{m}$ are
derived.  Finally, $\M$ is also corrected from its new fluxes and $\rho$ is
corrected from the new solution of $p$ according to the perfect gas law, before
moving forward to the next time step and returning to the sequence of equations
for $\rho$, $\M$ and $E$.

Our FV discretisation maintains a compact computational molecule for the
orthogonal component of the Laplacian terms, which corresponds to Rhie and Chow
interpolation \cite{Rhie&Chow:1982} in the pressure equation. Both the
transported fields in convection terms and the fluxes in $\M$ are interpolated
using limiters from the MUSCL total variation diminishing (TVD) scheme
\cite{VanLeer:1979,Hirsch:1990} with identical limiters used in all convection
terms (for $\rho$, $\M$ and $E$) to maintain numerical consistency. The temporal
derivative is discretised using a two-time-level Euler implicit scheme.

We calculated shocks of Mach 1.2, 1.7, 2.2, 2.84, 3.4, 4, 5, 6, 7, 8, 9, 10 and
11 in order to provide a reasonable distribution of solution points for
subsequent comparison with results from experiment. Mach 2.84 was specifically
chosen to coincide with shock profile data communicated to us privately
\cite{Torecki&Walenta:1993}. Similarly, Mach 8 was chosen to coincide with the
published shock profile data of Steinhilper~\cite{Steinhilper:1972}, and Mach 9
coincides with a published profile of Alsmeyer~\cite{Alsmeyer:1976}.

A solution domain of 33$\lambda_{M1}$ was used in all simulations --- wide
enough to contain the entire shock structure comfortably.  Our initial results
were obtained using the Navier-Stokes equations with the control viscosity
exponent $s = 0.72$.  The results for $\rho$ and $T$ converged on a mesh of 800
cells to within 1\% of the solution extrapolated to an infinitely small mesh
size.  The results we present in this paper were produced with a mesh of 2000
cells, corresponding to a mesh size of $\sim 0.017\lambda_{M1}$. Numerical
solutions were executed until they converged to steady-state, at which point the
residuals of all equations had fallen 5 orders of magnitude from their initial
level.

\section{Volume diffusivity}
\subsection{Unphysical behaviour when $\alpha_{v} \equiv \alpha$}
\label{sec:volumeDiffusivity}
\begin{figure}[ht]
    \centering
    \input{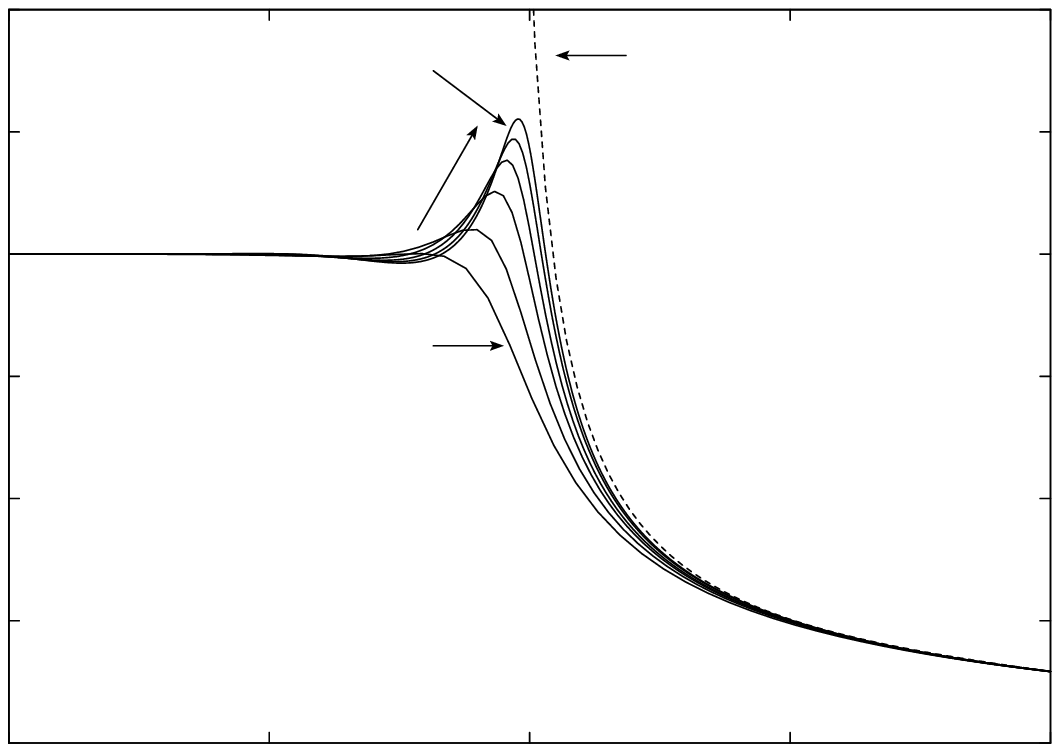}
    \caption{Mach number profiles for decreasing cell sizes, assuming $\alpha_{v} \equiv
    \alpha$.}
    \label{fig:MachOvershoot}
\end{figure}

We followed our initial Navier-Stokes simulations with preliminary simulations
using the Brenner-Navier-Stokes equations.  These simply used a volume
diffusivity $\alpha_{v} \equiv \alpha$, which was Brenner's original suggestion
(discussed in section~\ref{sec:modifiedNavierStokes} above).  However, our
simulations do not reach a converged solution with decreasing cell (or mesh)
size: at the upstream edge of the temperature profile a small undershoot
develops at a cell size of $~0.06\lambda_{M1}$ that increases in magnitude with
decreasing cell size.  The problem is present in profiles of all solution
variables but is best illustrated in a plot of Mach number, as shown in
fig.~\ref{fig:MachOvershoot}.  At best, the level of overshoot at the smallest
cell sizes seems unphysical; worse is that the overshoot may tend to infinity
as the mesh is further refined.

There is little doubt that the overshoot in Mach number is a
consequence of Brenner's modification.  In subsequent tests we were able to
attribute the presence of the overshoot to the additional term in the momentum
flux, but not to the additional term in the energy flux.  We therefore postulate
that the overshoot might be caused by an inappropriately large volume
diffusivity, particularly since it exceeds the diffusivity associated with the
remaining terms in the model, the kinematic viscosity $\nu  = \mu/\rho$, by a
factor $\alpha^{\star} = \alpha_{v}/\nu = {\Pr}^{-1} = 1.5$.

Our search for an alternative value of $\alpha_{v}$ began by relating the
physical process of volume diffusion more closely to mass diffusion, rather than
thermal diffusion.  However, the process of diffusion of mass within a single
component fluid, or {\em self-diffusion}, occurs at a similar rate to thermal
diffusion, with theoretical self-diffusivity coefficients $D_{m}$ of $1.200\nu$
for hard-sphere molecules and $1.534\nu$ for Maxwellian molecules
\cite{Kennard:1938}. The unphysical overshoot in Mach number remains for both
values of self-diffusion coefficient.  The overshoot is less pronounced when using
the lower value but is still increasing with descreasing cell size even at the
smallest cell sizes we tested ($~0.008\lambda_{M1}$).
Again, the amount of overshoot is unphysical and may tend to infinity as the
mesh is further refined.  Only when we reduce the
volume diffusivity coefficient to $\alpha_{v} = \nu$
(i.e.\ when we set $\alpha^{\star} = 1$) does the overshoot disappear.

\subsection{Investigation of the unphysical behaviour}
\label{sec:unphysicalBehaviour}
It is known that some forms of extended hydrodynamic equations are unstable in
time to periodic spatial disturbances with wavelengths shorter than a critical
length that is typically of the order of one mean free path
\cite{Zhong&al:1991}. Such instabilities appear in numerical simulations when
the mesh is sufficiently fine to resolve wavelengths shorter than the critical
length, i.e.\ when the numerical cell length is below this critical length.  The
appearance of an overshoot below a critical level of mesh refinement in our
simulations may indicate a similar instability, although the overshoot does
appear at a particularly short cell length and the solutions do converge to a
steady state and so do not `blow up' in time.

Similarly, some forms of extended hydrodynamic equations, which may be stable in
time, are actually unstable in space to periodic temporal disturbances
\cite{Struchtrup&Torrilhon:2003,Torrilhon&Struchtrup:2004,Struchtrup:2005}.
Again, it is unclear that such an instability would produce the overshoot
behaviour we witnessed in our preliminary calculations.  Nevertheless, here we
undertake both temporal and spatial stability analyses of the
Brenner-Navier-Stokes equations in order to investigate the source of unphysical
behaviour.

Following the procedures described in
\cite{Zhong&al:1991,Struchtrup&Torrilhon:2003} the governing equations from
section~\ref{sec:modifiedNavierStokes} are first linearised in 1-dimension to
produce the following non-dimensionalised perturbation equations:
\begin{equation}
    \frac{\partial \phi}{\partial t^{\prime}}
  +
    \begin{bmatrix}
        \ 0\  & \ 1\ & \ 0\ \\
        1 & 0 & 1 \\
        0 & \frac{2}{3} & 0
    \end{bmatrix}
    \frac{\partial \phi}{\partial x^{\prime}}
  + \frac{\partial}{\partial x^{\prime}}
    \left\{
        \begin{array}{c}
            0 \\ \sigma^{\prime} \\ \frac{2}{3} q^{\prime}
        \end{array}
    \right\}
  =
    0 ,
    \label{eq:pertubation}
\end{equation}
where $t^{\prime}$ and $x^{\prime}$ are nondimensionalised time and distance,
respectively, $\phi = \{\rho^{\prime}\ u^{\prime}\ T^{\prime}\}^{\mathrm{T}}$
is the vector of nondimensionalised density, flow speed and temperature, and
$\sigma^{\prime}$ and $q^{\prime}$ are nondimensionalised momentum and heat
fluxes respectively.  From eqs.~(\ref{eq:NewtonianFluidVolumeVelocity}) and
(\ref{eq:internalEnergyFluxDensityFinal}), these momentum and energy fluxes are,
respectively,
\begin{equation}
    \sigma^{\prime}
  =
  - \frac{4}{3} \frac{\partial u^{\prime}}{\partial x^{\prime}}
  - \frac{4}{3} \alpha^{\star}
    \frac{\partial^{2} \rho^{\prime}}{\partial x^{\prime 2}} ,
\end{equation}
and
\begin{equation}
    q^{\prime}
  =
  - \frac{15}{4} \frac{\partial T^{\prime}}{\partial x^{\prime}}
  - \alpha^{\star} \frac{\partial \rho^{\prime}}{\partial x^{\prime}} .
\end{equation}
We assume a solution to eq.~(\ref{eq:pertubation})
%
of the form
\begin{equation}
    \phi
  =
    \tilde{\phi}
    \exp\left\{\im(\omega t^{\prime} - k x^{\prime})\right\} ,
    \label{eq:perturbationSolution}
\end{equation}
where $\tilde{\phi}$ is the amplitude of the wave, $\omega$ is its
frequency and $k$ its propagation constant. Equations~(\ref{eq:pertubation}) to
(\ref{eq:perturbationSolution}) can be combined to produce a set of linear
algebraic equations of the form
\begin{equation}
    \mathcal{A}(\omega,k) \tilde{\phi} = 0 ,
    \label{eq:linearAlgebraicPerurbation}
\end{equation}
for which non-trivial solutions require
\begin{equation}
    \det[\mathcal{A}(\omega,k)] = 0 .
    \label{eq:linearAlgebraicPerurbationDeterminant}
\end{equation}
%

For the Brenner-Navier-Stokes equations,
eq.~(\ref{eq:linearAlgebraicPerurbationDeterminant}) yields the following
characteristic equation:
\begin{equation}
    6\im  \omega^{3}
  + 23 k^{2} \omega^{2}
  - [10 k^{2} + (20 + 8 \alpha^{\star})k^{4}]\im \omega
  - [(15 - 4\alpha^{\star}) k^{4} + 20\alpha^{\star}k^{6}]
  = 0 .
    \label{eq:BNScharacteristics}
\end{equation}
If a disturbance in space is considered as an initial-value problem, $k$ is real and $\omega = \omega_{r} + \im\omega_{i}$ is complex.  The form of eq.~(\ref{eq:perturbationSolution}) indicates that stability then requires $\omega_{i} \ge 0$.  If a disturbance in time is considered as a problem of signalling from the boundary, $\omega$ is real and $k = k_{r} + \im k_{i}$ is complex. For a wave travelling in the positive $x$ direction, $k_{r} > 0$, and stability then requires that $k_{i} < 0$.  For a wave travelling in the negative $x$ direction, the converse is true: $k_{r} < 0$ and stability requires $k_{i} > 0$.

\begin{figure}[ht]
    \centering
    \input{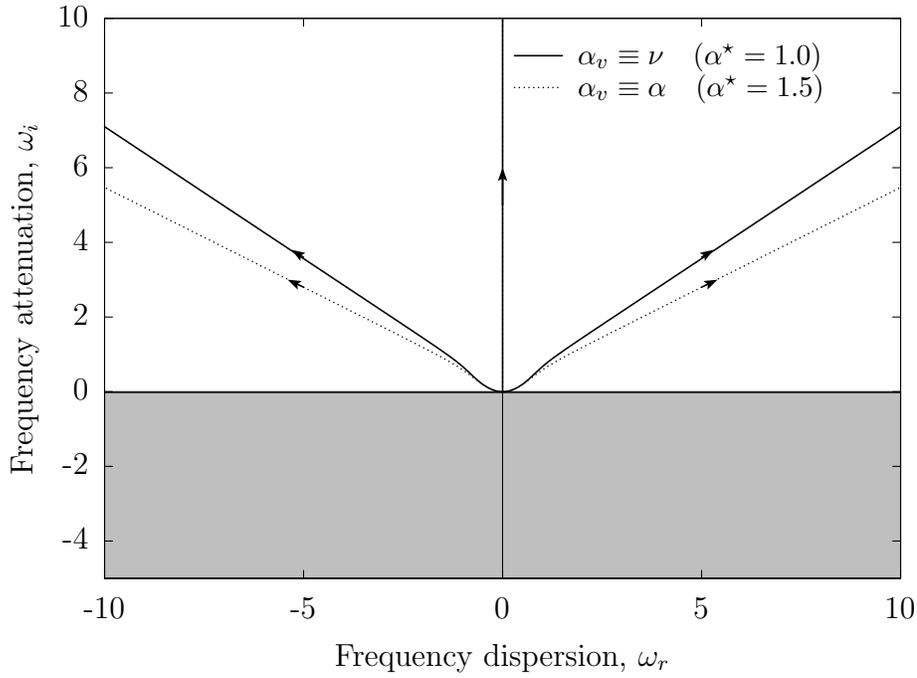}
    \caption{Temporal stability of Brenner-Navier-Stokes
equations (arrows indicate direction of increasing $k$; grey shaded area
indicates region of instability).}
    \label{fig:TemporalStabilityBrenner}
\end{figure}

We examine temporal stability by solving eq.~(\ref{eq:BNScharacteristics})
numerically for $\omega$ for values of $k$ in the range $0 \le k < \infty$.
Trajectories of $\omega$ are plotted in the complex plane in
fig.~\ref{fig:TemporalStabilityBrenner}. Two sets of trajectories are plotted:
those for $\alpha_{v} \equiv \alpha$  (corresponding to $\alpha^{\star} =
{\Pr}^{-1} = 1.5$) and those for $\alpha_{v} \equiv \nu$ (for which
$\alpha^{\star} = 1.0$).  In both cases the trajectories all lie within the
region $\omega_{i} \ge 0$, indicating stability for all $k$.  This confirms, as
expected, that the observed overshoot is not caused by temporal instability.

\begin{figure}[ht]
    \centering
    \input{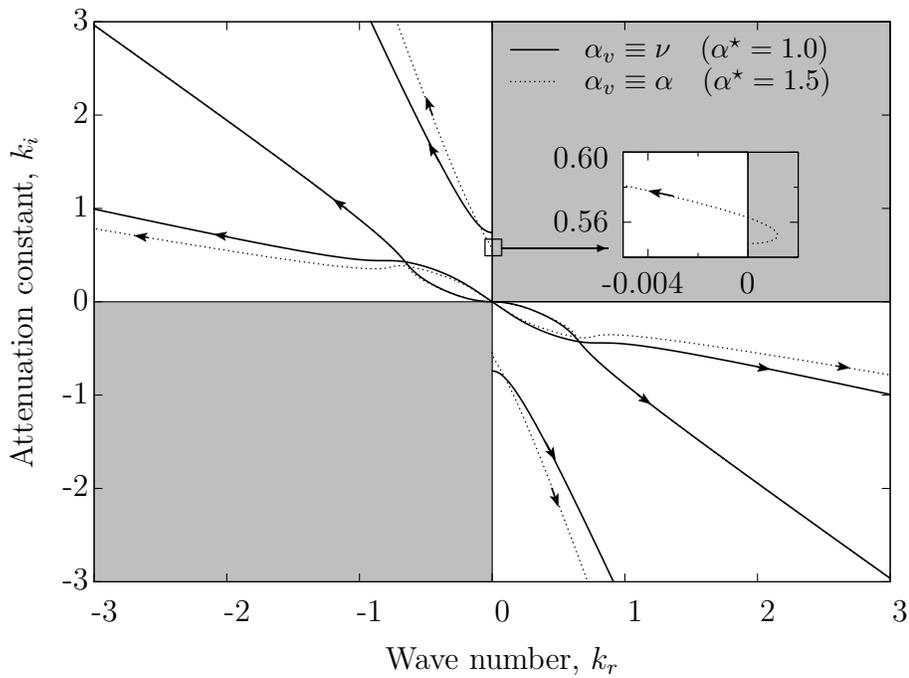}
    \caption{Spatial stability of Brenner-Navier-Stokes equations (arrows
indicate direction of increasing $\omega$; grey shaded areas
indicate regions of instability).}
    \label{fig:SpatialStabilityBrenner}
\end{figure}

We then turn to examine spatial stability by solving
eq.~(\ref{eq:BNScharacteristics}) numerically for $k$ for values of $\omega$ in
the range $0 \le \omega < \infty$.  Trajectories of $k$ are plotted in the
complex plane in fig.~\ref{fig:SpatialStabilityBrenner} for both $\alpha_{v}
\equiv \nu$ and $\alpha_{v} \equiv \alpha$.  When $\alpha_{v} \equiv \nu$ the
trajectories do not violate the stability condition.  However, when $\alpha_{v}
\equiv \alpha$, the inset graph shows one trajectory start from
$\omega = 0$,  $k_{r} = 0$ at $k_{i} \approx 0.55$,
enter the unstable region $\{k_{r} > 0, k_{i} > 0\}$,
and exit into the stable region $\{k_{r} < 0, k_{i} > 0\}$ by crossing the $k_{r} = 0$ axis at $k_{i} \approx 0.56$.  Thus, the stability condition is clearly violated for small $\omega$.

Subsequently we examined trajectories for a number of different
$\alpha^{\star}$ and we found that the equations are unstable for
$\alpha^{\star} \gtrsim 1.45$, which suggests a potential problem for some intuitive
choices of $\alpha_{v}$, such as $\alpha$ and $D_{m}$.  However, this result
does not really explain the cause of the overshoot in Mach number, since the overshoot only disappears when $\alpha^{\star}$ falls to unity, i.e.\ considerably lower than 1.45.

\begin{figure}[ht]
    \centering
    \input{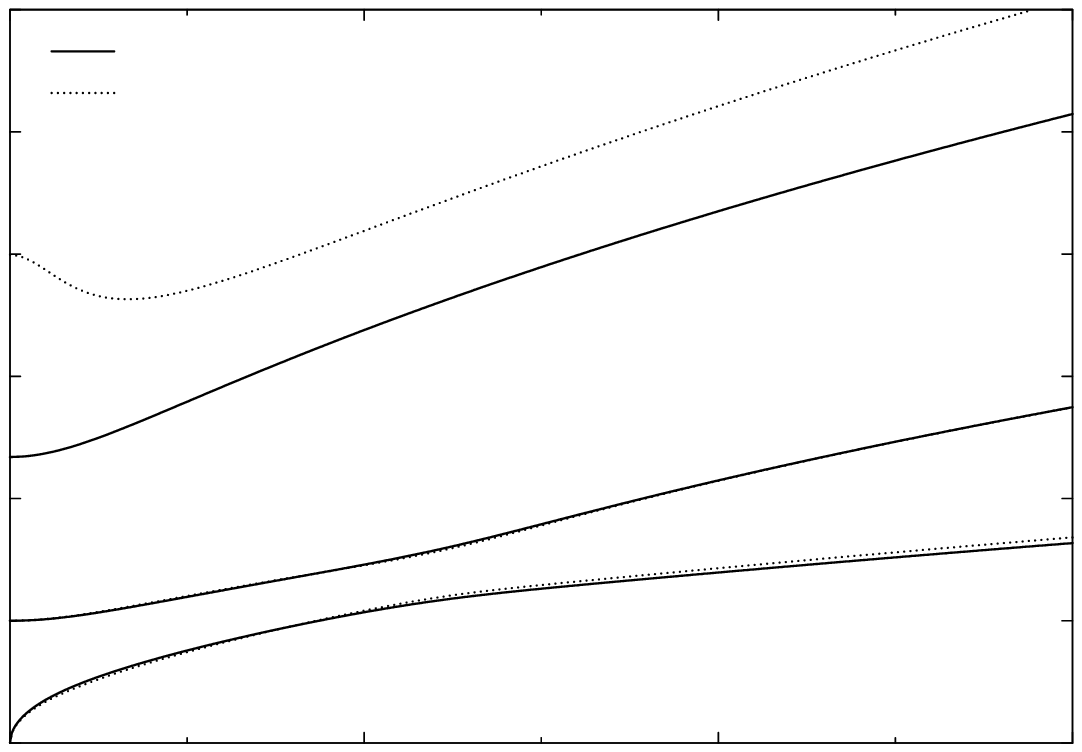}
    \caption{Normalised phase velocity for the Brenner-Navier-Stokes equations.}
    \label{fig:PhaseVelocityBrenner}
\end{figure}

Unphysical behaviour can also be observed by examining the phase velocity:
\begin{equation}
    v_{ph} = \frac{\omega}{k_{r}(\omega)}.
    \label{phaseVelocity}
\end{equation}
Figure~\ref{fig:PhaseVelocityBrenner} shows the phase velocity,
normalised by the speed of sound in the nondimensionalised form in the
perturbation equations ($c_{0} = \sqrt{\gamma}$), for $\alpha^{\star} = 1.0$ and
$\alpha^{\star} = 1.2$.  For both $\alpha^{\star}$, results for mode 1 are
superimposed and correspond to the propagation of sound. The
mode 2 results correspond to the diffusive transport of heat and results for
both $\alpha^{\star}$ are very similar.  However, there is a marked difference
between results for the two cases for mode 3, relating to higher-order diffusive transport.  The $\alpha^{\star} = 1.0$
results are similar to those of
other extended hydrodynamic models, e.g.\ the super-Burnett
equations~\cite{Struchtrup&Torrilhon:2003}, beginning at a moderate speed,
$v_{ph}/c_{0} = 2.34$ at $\omega = 0$, before increasing steadily with
increasing $\omega$.  The $\alpha^{\star} = 1.2$ results are, however, unusual:
the phase velocity at $\omega = 0$, i.e.\ $v_{ph}/c_{0} = 3.99$, is high in
comparison to other hydrodynamic models~\cite{Struchtrup&Torrilhon:2003}. The
phase velocity also decreases initially with increasing $\omega$, before passing
through a minimum and increasing thereafter.  The high initial phase velocity
seems anomalous, and rises to extraordinary levels for higher $\alpha^{\star}$,
e.g.\ if $\alpha^{\star} = 1.5$, $v_{ph}/c_{0} = 190.1$ at $\omega = 0$.
The initial decrease
in phase velocity with increasing $\omega$ may allow low frequency waves
upstream of the shock to overtake slower, higher frequency waves within the
shock, creating counter-dispersion at the upstream end of the shock.  The
initial decrease in phase velocity disappears only when $\alpha^{\star}$ falls
below $\sim 1.11$, a level quite close to that at which we find the unphysical
overshoot disappears.

To summarise, our results show unphysical behaviour for the
Brenner-Navier-Stokes equations when $\alpha^{\star} = 1.5$. A spatial stability
test confirms the equations are unstable to temporal disturbances when
$\alpha^{\star} \gtrsim 1.45$.  Plots of phase velocity raise further questions
about the physical nature of the solutions when $\alpha^{\star} \gtrsim 1.11$.
Taken together, this casts doubt both on Brenner's proposed $\alpha_{v} \equiv
\alpha$ and on the apparently natural choice of $\alpha_{v} \equiv D_{m}$.  The
overshoot in Mach number disappears when $\alpha^{\star} = 1.0$, i.e.
$\alpha_{v} \equiv \nu$.  We therefore adopt this model for $\alpha_{v}$ in the
Brenner-Navier-Stokes equations for the remainder of this paper.

\section{Results and comparison with experiment}
\label{sec:results}
\subsection{Shock profiles}
\label{sec:shockProfiles}
We prefer, where possible, to compare results with actual experiments rather
than DSMC simulations, since the latter requires certain assumptions relating to
the form of the intermolecular force law.  We therefore first make comparison
with actual measured data for the variation of density through the shock layer
\cite{Steinhilper:1972,Torecki&Walenta:1993,Alsmeyer:1976}.

\begin{figure}[ht]
    \centering
    \input{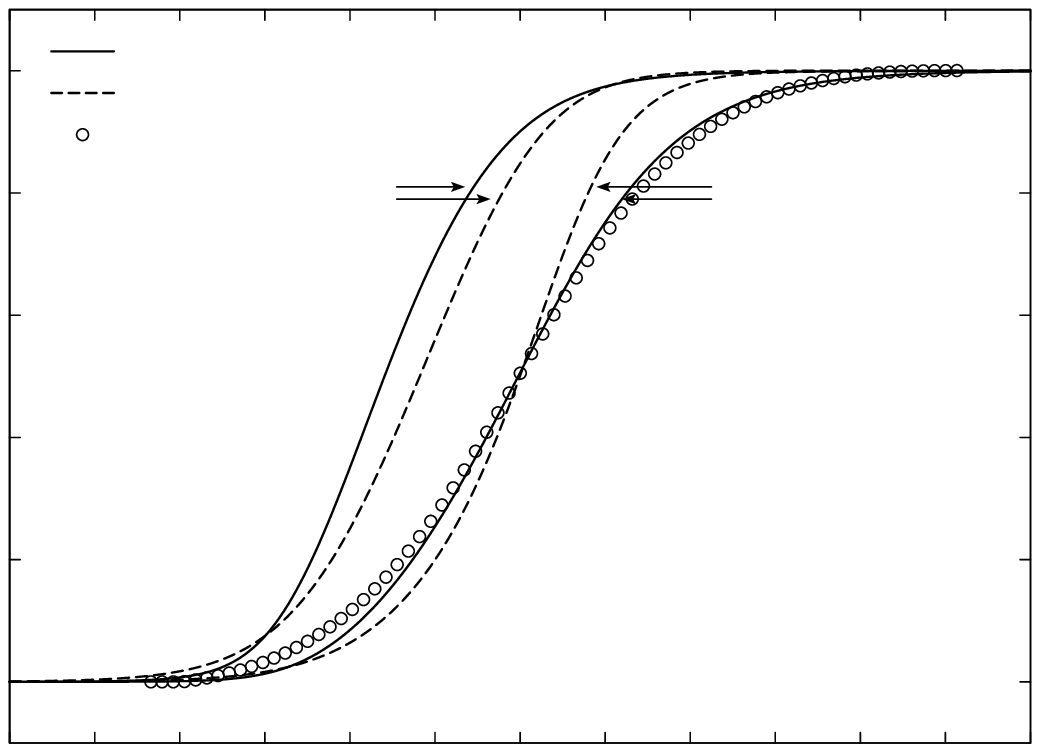}
    \caption{Simulated and experimental profiles of a Mach 2.84 stationary
             shock; $s = 0.72$.}
    \label{fig:Mach284profile}
\end{figure}

Figure~\ref{fig:Mach284profile} shows the normalised variation of density and
temperature, $\rho^{\star}$ and $T^{\star}$ respectively, through an argon gas
shock of Mach 2.84 calculated using the Navier-Stokes and Brenner-Navier-Stokes
equations with $s = 0.72$. The experimental density profile
\cite{Torecki&Walenta:1993} is also shown. It is clear that the shock layer
predicted by the conventional Navier-Stokes equations is too thin, whereas the
Brenner-Navier-Stokes equations produce good agreement with the experimental
data. The main region of disparity is upstream of the shock layer (left hand
side in the figure) where the experimental data trails out and is flatter than
the prediction.

\begin{figure}[ht]
    \centering
    \input{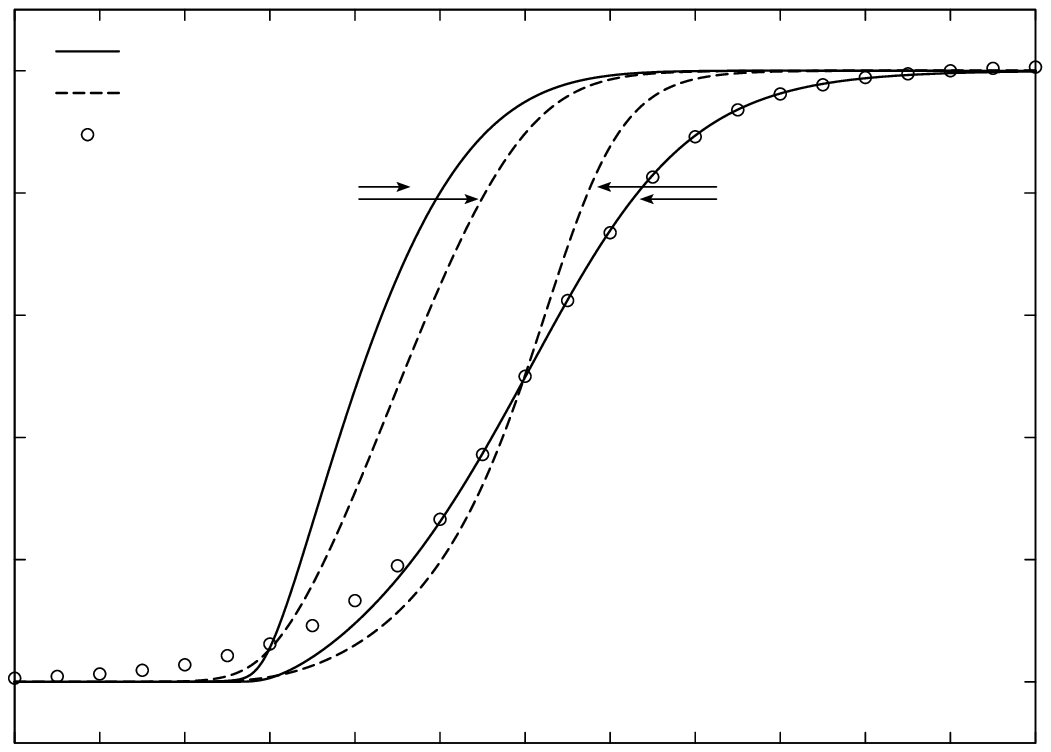}
    \caption{Simulated and experimental profiles of a Mach 8.0 stationary shock;
$s = 0.72$.}
    \label{fig:Mach8profile}
\end{figure}

Similarly, fig.~\ref{fig:Mach8profile} shows the predicted profiles for a Mach
8.0 shock compared with experimental density data \cite{Steinhilper:1972}.
Again, the Navier-Stokes equations produce a shock profile which is too thin
when compared with experiment. However, the Brenner-Navier-Stokes equations
produce excellent agreement in the central and downstream shock regions
($\rho^{\star} > 0.2$), only in the upstream region is the predicted profile
sharper than the experimental data shows --- just as in the Mach 2.84 case.

\begin{figure}[ht]
    \centering
    \input{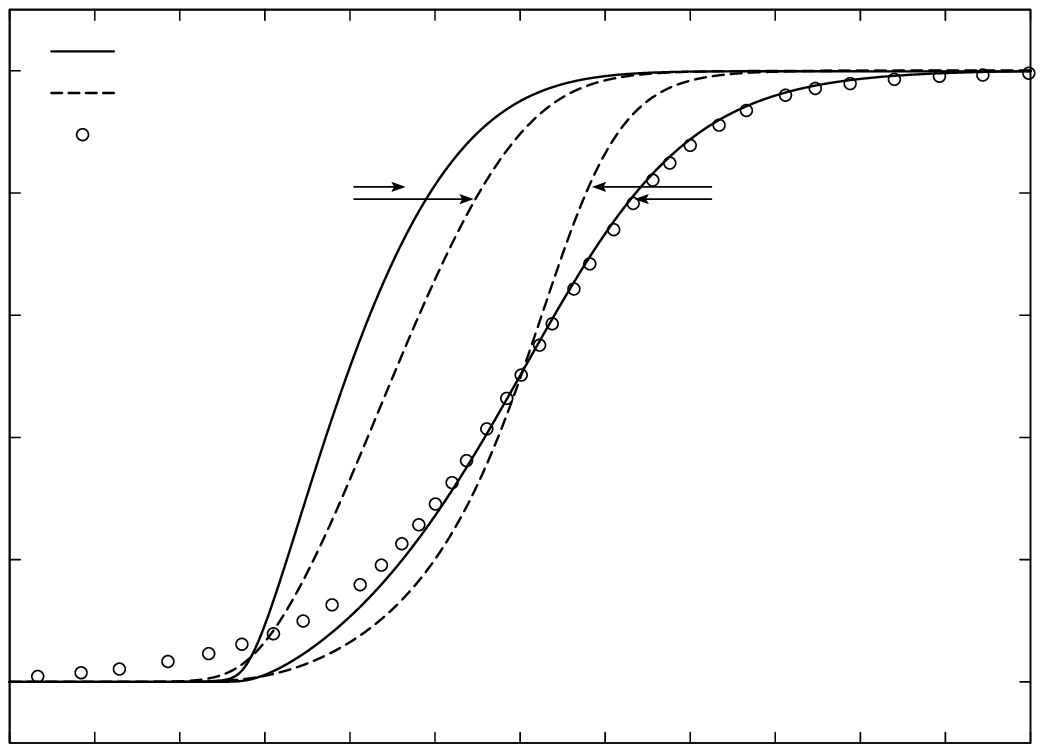}
    \caption{Simulated and experimental profiles of a Mach 9.0 stationary shock;
$s = 0.72$.}
    \label{fig:Mach9profile}
\end{figure}

Figure~\ref{fig:Mach9profile} shows results for a Mach 9.0
shock~\cite{Alsmeyer:1976}.  In this case our observations are very similar to
those we make about the Mach 8.0 shock profile, above; this figure is included
here for completeness.

\subsection{Inverse density thickness}
\label{sec:inverseDensityThickenss}
Apart from direct comparison of calculated and experimental shock profiles,
there are other shock parameters for which experimental and/or independent
numerical data is available.  The principal parameter is the non-dimensional
shock inverse density thickness, defined as:
\begin{equation}
L_{\rho}^{-1} = \frac{\lambda_{M1}}{\rho_{2} - \rho_{1}} |\grad\rho|_{\max}.
\label{eq:inverseDensityThickness}
\end{equation}
In the absence of an {\em a priori} characteristic length scale in an unconfined
flow, the definition of $\Kn$ requires a characteristic dimension of a flow
structure, in this case the actual thickness of the shock layer itself.
Therefore $L_{\rho}^{-1}$ has the interesting feature that it represents $\Kn$
for the shock structure case\footnote{While this identification then indicates,
as we see below, that shocks generally have such a high overall $\Kn$ that any
hydrodynamic model should fail, we can still assess extended hydrodynamic
equations for their usefulness as engineering models.}.

\begin{figure}[ht]
    \centering
    \input{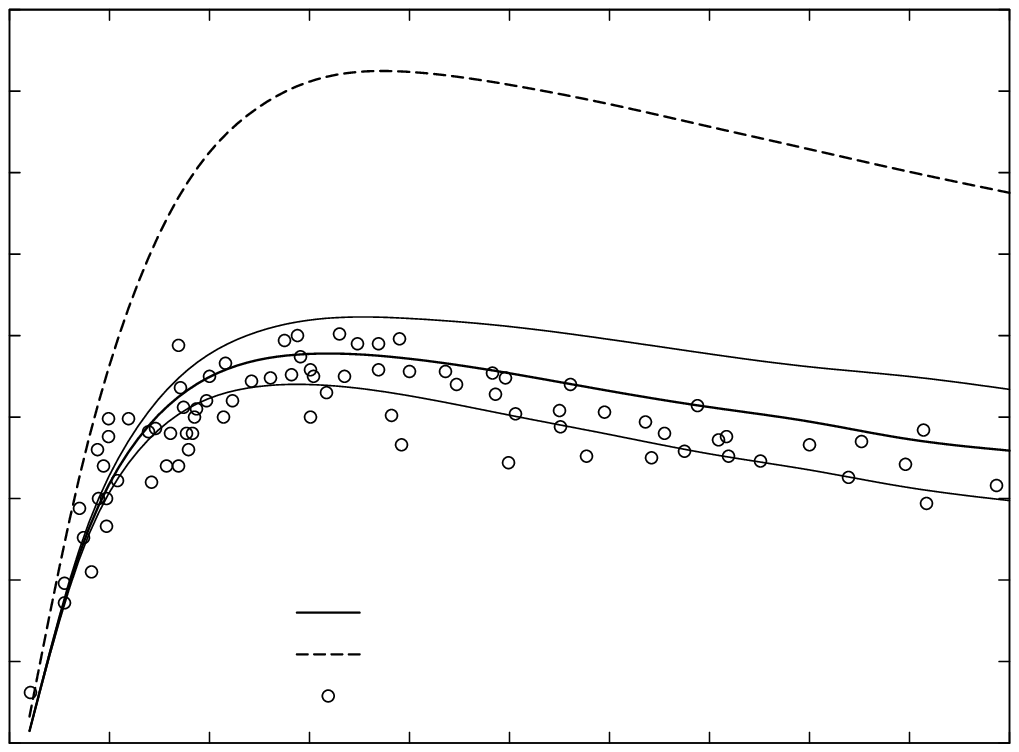}
    \caption{Simulated and experimental inverse density thickness ($L_{\rho}^{-1}$) data, versus shock Mach number; for various values of the exponent $s$.}
    \label{fig:MachVsInvShockThickness}
\end{figure}

Alsmeyer \cite{Alsmeyer:1976} reported the most comprehensive collection of
experimental shock data, comprising previously-published work as well as his own
new results. Figure~\ref{fig:MachVsInvShockThickness} shows $L_{\rho}^{-1}$ for
argon shocks up to Mach 11, with experimental data collated from a number of
sources \cite{Steinhilper:1972,Alsmeyer:1976,Torecki&Walenta:1993}. The
Navier-Stokes equations, with $s=0.72$, predict shocks of approximately half the
measured thicknesses over the entire Mach number range.  As $L_{\rho}^{-1}$
indicates $\Kn$, this poor agreement is expected because we can see that $\Kn
\sim 0.2$ -- $0.25$ over most of this Mach number range, so the Navier-Stokes
equations are beyond their effective range of application. However, the results
from the Brenner-Navier-Stokes equations closely match experiment, with moderate
sensitivity to the choice of viscosity-temperature exponent: using 0.72 (the
control value) and 0.76 (our best-fit value for temperatures equivalent to
shocks up to Mach 12.5), the results fall within the limits of experimental
scatter; using $s = 0.68$, the results stray slightly outside the scatter of
experimental data just before they reach the exponent's limit of applicability
at Mach 4.3.  With the results for 0.72 at the higher end of the experimental
scatter, and results for 0.76 at the lower end, we estimate that an exponent of
$s \approx 0.74$ would produce the best agreement with the experimental results.

\subsection{Density asymmetry quotient}
\label{sec:densityAsymmetryQuotient}
Agreement of predicted and experimental shock inverse density thicknesses is not
the only measure of the success of a new model. As $L_{\rho}^{-1}$ depends on
the density gradient at the profile midpoint alone, it does not express anything
about the overall shape of the profile.
If $L_{\rho}^{-1}$ is used as the sole parameter to describe the shock it could
be concluded that the Brenner-Navier-Stokes equations tested here, and many
other models previously published, have excellent predictive capability.
However, the shock profiles in figs.~\ref{fig:Mach284profile} and
\ref{fig:Mach8profile} show there are differences between simulation and
experiment, in particular relating to the flatter region upstream of the profile
that is observed experimentally.

\begin{figure}[ht]
    \centering
    \input{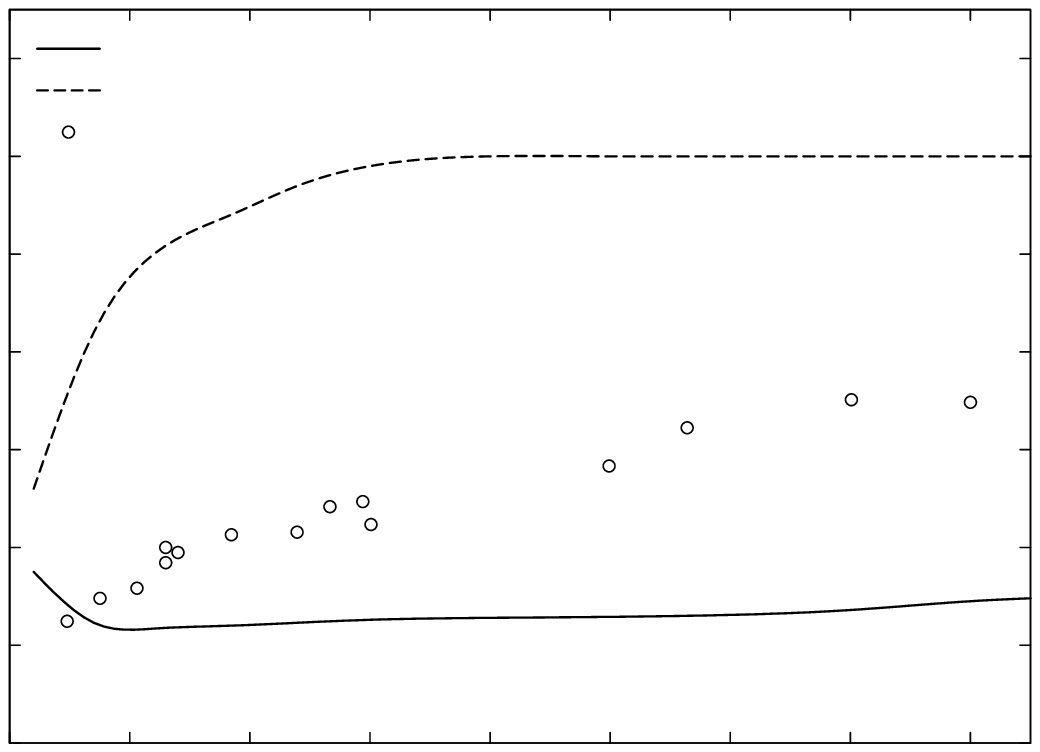}
    \caption{Simulated and experimental asymmetry quotient ($Q_{\rho}$) data, versus   shock Mach number; $s = 0.72$.}
    \label{fig:MachVsAsymmetryQuotient}
\end{figure}

Therefore, a second parameter which should be used to describe the shock
profile, and for which experimental data is available, is the density asymmetry
quotient $Q_{\rho}$.  This is a measure of how skewed the shock density profile
is relative to its midpoint. It is defined for a 1-dimensional profile of
normalised density, $\rho^{\star}$, centred at $\rho^{\star} = 0.5$ on $x = 0$,
as
\begin{equation}
    Q_{\rho}
  =
    \frac%
    {\int_{-\infty}^{0} \rho^{\star}(x)\,\text{d}x}
    {\int_{0}^{\infty} [1 - \rho^{\star}(x)]\,\text{d}x} .
\label{eq:asymmetryQuotient}
\end{equation}
A symmetric shock would consequently have $Q_{\rho} = 1$, but real shock waves
are not completely symmetrical about their midpoint.  First, their `bulk' form
is generally skewed somewhat towards the downstream.  Then, the aforementioned
flattened, diffusive region, that extends upstream of the shock profile, tends
to increase $Q_{\rho}$. Figure~\ref{fig:MachVsAsymmetryQuotient} shows
experimental data compiled by Alsmeyer \cite{Alsmeyer:1976} in which $Q_{\rho}$
increases fairly linearly from $\sim 0.9$ at around Mach 1.5, through unity at
around Mach 2.3, to $\sim 1.15$ at Mach 9. The bulk form therefore corresponds
to $Q_{\rho}\approx 0.9$ and the upstream flattened region accounts for a
further increase in $Q_{\rho}$, of up to 0.25 at Mach 9.

Results from the Navier-Stokes equations do not agree well with experimental
data: the bulk form is skewed towards the upstream side so that $Q_{\rho} > 1$
even at the lowest Mach numbers and the skewness further increases with Mach
number (apparently by sharpening of the profile downstream rather than
flattening upstream) so that by Mach 4, $Q_{\rho} \approx 1.4$, compared to
$\sim 1.03$ from experiment.

We find the Brenner-Navier-Stokes equations predict the bulk form of the profile
very well, predicting $Q_{\rho}\approx0.9$ at low Mach numbers.  As discussed in
section~\ref{sec:shockProfiles}, it does not capture the flattened region
upstream and so the departure from experimental data increases with Mach number.

\subsection{Temperature-density separation}
\label{sec:temperatureDensitySeparation}
The final shock structure parameter is the temperature-density separation,
$\delta_{T\rho}$, measured between the midpoints of the respective normalised
profiles. In a shock, the density rises from its upstream value to its
downstream value behind the temperature, due to the finite relaxation times for
momentum and energy transport; a good hydrodynamic model should resolve this
spatial lag accurately.  However, experimental data for this parameter is scarce
due to the difficulty in measuring temperature profiles, so independent DSMC
data \cite{Lumpkin&Chapman:1991} is usually taken for comparison.

\begin{figure}[ht]
    \centering
    \input{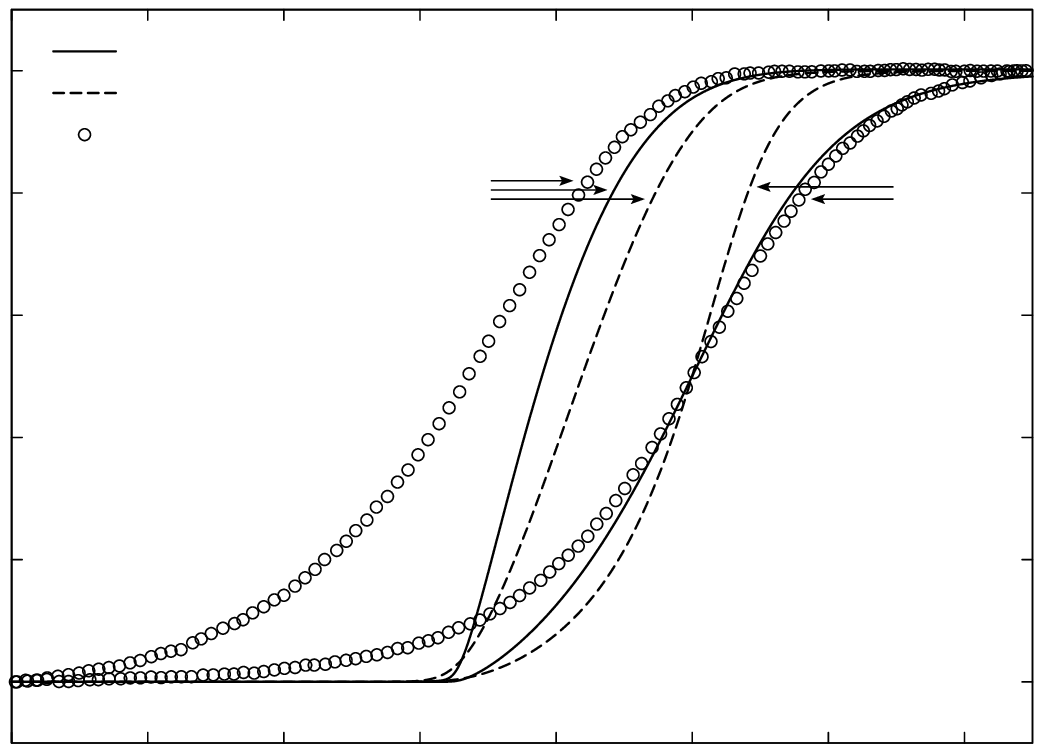}
    \caption{Simulated and DSMC profiles of a Mach 11 stationary
             shock; $s = 0.72$.}
    \label{fig:Mach11profile}
\end{figure}

Figure~\ref{fig:Mach11profile} shows the temperature and density profiles for a Mach 11 shock calculated using DSMC~\cite{Lumpkin&Chapman:1991}, and the Navier-Stokes and the Brenner-Navier-Stokes models. As in our earlier comparisons
in section~\ref{sec:shockProfiles}, the Brenner-Navier-Stokes equations produce
profiles that are much sharper in the upstream region of the shock than those
from DSMC results.

\begin{figure}[ht]
    \centering
    \input{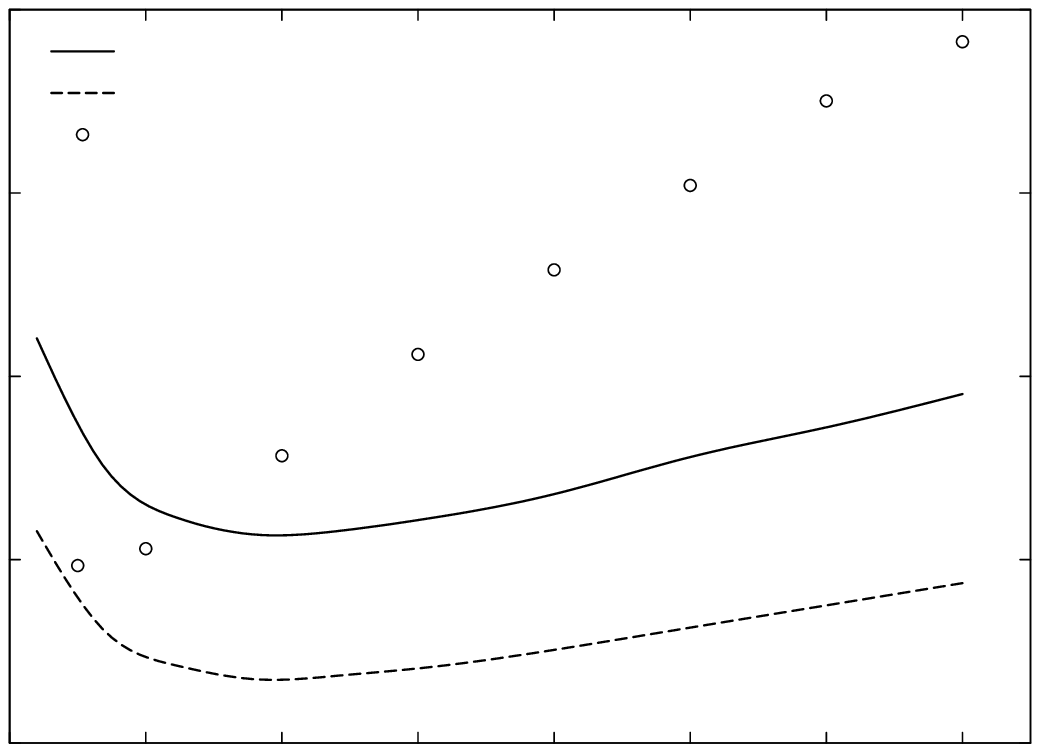}
    \caption{Simulated and independent DSMC temperature-density separation
    ($\delta_{T\rho}$) data, versus Mach number; $s = 0.72$.}
    \label{fig:MachVsDeltaTrho}
\end{figure}

Figure~\ref{fig:MachVsDeltaTrho} compares DSMC data for $\delta_{T\rho}$ over a range of shock Mach numbers with results from our simulations. The DSMC data show an increase in $\delta_{T\rho}$ from $\sim 1.5
\lambda_{M1}$ at Mach 1.5 to $\sim 2.9 \lambda_{M1}$ at Mach 8.  The
Navier-Stokes and Brenner-Navier-Stokes equations increasingly under-predict
$\delta_{T\rho}$ with increasing Mach number, although the Brenner-Navier-Stokes
equations generally perform a little better over the Mach number range.

\subsection{Very strong shocks}
\label{sec:verystrongshocks}
The inverse density thickness of extremely strong shocks is a useful additional
comparison for any proposed hydrodynamic model. Narasimha and Das
\cite{Narasimha&Das:1986} examined the solution of the Boltzmann equation for an
infinitely strong shock (a more recent treatment is in
\cite{Cercignani&Al:1999}), modelling the upstream flow as a molecular beam
with a distribution function of the form $f(x =
-\infty)~=~n_{1}\delta(\vec{u}_{1})$, where $\delta$ is the Kronecker delta
function. The shock layer may then be treated as a device for converting this
beam function into a downstream Maxwellian distribution function. The
distribution function in the shock layer can be expressed as a linear
combination of the two extremal distribution functions, a method similar to the
bimodal method of \cite{Muckenfuss:1962}.

Using an expansion parameter that measures the departure of the distribution
function in the shock wave from that outlined in the previous paragraph, an
infinite series of ordinary differential equations is obtained for the shock
thickness. This series rapidly converges, and a solution of the first seven
equations of the set yields a predicted shock thickness of $6.7 {\lambda}_{M2}$
(which is written in terms of the downstream Maxwellian mean free path version
of eq.~\ref{eq:meanFreePathMaxwell}).  When this is converted into the
$L_{\rho}^{-1}$  of eq.~(\ref{eq:inverseDensityThickness}), the inverse density
thickness for a shock with a downstream $\lambda_{M2}$ equivalent to that for a
shock of Mach 100 is predicted to be 0.076.

Our calculations for shocks of Mach 100 give $L_{\rho}^{-1} = 0.156$ for the
Navier-Stokes equations with $s = 0.72$. The results with the
Brenner-Navier-Stokes equations are $L_{\rho}^{-1} = 0.091$ for $s = 0.72$, and
$L_{\rho}^{-1} = 0.066$ for $s = 0.76$.  These values for $L_{\rho}^{-1}$
straddle the solution from the molecular beam analysis.  Further simulations
with successive adjustments to $s$ gave a precise match in $L_{\rho}^{-1}$ for
$s = 0.742$, which is in agreement with the exponent estimated in
section~\ref{sec:inverseDensityThickenss} to produce the best agreement with
experiment over the range Mach 1--11.

\section{Discussion and conclusions}
\label{sec:discussion}
The Navier-Stokes equations are robust and accurate over a wide range of $\Kn$
--- surprisingly so, given some of the relatively narrow axioms on which they
depend (i.e.\ the continuum-fluid and local-equilibrium requirements). Such a
good fluids engineering model is difficult to relinquish, even when flow systems
well beyond its range of applicability are considered
\cite{Herwig&Hausner:2003}. However, it is clear from the results in
section~\ref{sec:results} that the Navier-Stokes equations fail in nearly every
respect in predicting correct shock structures above about Mach 2 (or,
equivalently, for intermediate-$\Kn$ flows).

While it is important not to draw strong conclusions based on just one test
case, our results are generally encouraging for the Brenner-Navier-Stokes
equations.  This modified model is significantly better at reproducing the
trends in the experimental and DSMC data, and in the case of the inverse density
thickness delivers an excellent match.  It is only the more detailed features of
the shock profile that Brenner's model seems unable to reproduce.

First, it does not predict the flattened upstream region, as discussed in
section~\ref{sec:densityAsymmetryQuotient}.  In this regard, a major advantage
of DSMC as a technique for simulating intermediate-$\Kn$ flows in general
is its ability to produce non-Maxwellian velocity
distributions, that may also differ in directions
parallel and perpendicular to the flow. It is not clear that hydrodynamic
models will be able to properly incorporate this physics, and certainly the
problems the Brenner-Navier-Stokes equations have in capturing the upstream
shock region properly is related to the fact that in this region the velocity
distribution
function is a non-Maxwellian combination of fast, cold upstream molecules
and slower, hot molecules that have diffused from downstream regions.

The second feature is also related to this distribution function problem:
bi-modal methods (see,
e.g., \cite{Cercignani&Al:1999}) for a hard-sphere gas predict a downstream
temperature overshoot of around 1\%, which is confirmed by careful DSMC
simulations. There are no downstream overshoots predicted in any of the
Brenner-Navier-Stokes shock simulations.

While some of these features can be obtained using certain extended hydrodynamic
models that are formally $O(\Kn^{2})$ (see, e.g.,
\cite{Lumpkin&Chapman:1991,Reese:1993}), this is at a cost: there are known
problems of physical stability, and the numerical implementation is difficult
due to the large number of additional non-linear and high order derivatives.
The Brenner modification does not suffer so much from these problems, having
only a single additional term in each of the momentum and energy conservation
equations.  That the adoption of these terms provides a substantial improvement
in predicted results raises the question of whether this model can compensate,
in part at least, for increased non-local-equilibrium in the gas, or whether
this agreement is coincidental.

Brenner proposed his modifications partly to understand how some effects that
are traditionally thought of as becoming important only in a flow approaching
the intermediate-$\Kn$ range, e.g.\ slip at solid bounding surfaces, can be
encompassed in a model which still retains its essential $O(\Kn)$
character~\cite{Brenner:2005a,Brenner:2005b}.  He shows that the form (if not
the exact coefficients, except for a particular molecular model) of two
particular terms that appear in the Burnett constitutive model for $\vStress$
are in fact encompassed by the additional term in $\rho$ in his modified
Newtonian $\vStress$ of eqn.~(\ref{eq:NewtonianFluidVolumeVelocity}). While all
the stress terms in the Burnett equations are formally $O(\Kn^{2})$, under some
circumstances these two terms can be of similar magnitude to those of $O(\Kn)$
i.e.\ the same order of accuracy as the Navier-Stokes equations. If the issue of
the correct model for the volume diffusivity, $\alpha_{v}$, can be resolved then the Brenner-Navier-Stokes equations may provide a simple
alternative to the family of extended hydrodynamic models
that includes those of Burnett, Grad, etc., producing reasonably accurate solutions
of intermediate-$\Kn$ flows at a modest computational cost. While it is known
that the classical Burnett equations do not satisfy the second law of thermodynamics,
truncated or extended forms of the equations can be constructed that do \cite{Lumpkin&Chapman:1991,Zhong&al:1991,Reese:1993}. The fact that the
Brenner-Navier-Stokes equations are less prone to both numerical instability and
unphysical solution may indicate that thermodynamic consistency is less of a
problem with these equations than with more complex extended hydrodynamics models.

We recognise that it is not reasonable to rely on one benchmark case to decide the
value of the
Brenner-Navier-Stokes equations, or any other extended hydrodynamic model (or
its associated boundary conditions). Equation~(\ref{eq:Knudsen}) indicates there
are three distinct categories of near-equilibrium flows:
\begin{enumerate}
\item[{\bf A.}] $\Ma = O(1)$, $\Re\rightarrow\infty$, typical of super- and
hypersonic flows;
\item[{\bf B.}] $\Ma\rightarrow 0$, $\Re = O(1)$, typical of flows in micro- or
nano-systems;
\item[{\bf C.}] $\Ma\rightarrow 0$, $\Re\rightarrow\infty$, typical of
incompressible turbulent boundary layer flows.
\end{enumerate}
As $\Kn$ vanishes more quickly for flows in category C than in categories A and
B, departures from local equilibrium in category C flows are not as significant
as in categories A and B flows. This paper has addressed a category A flow in
which the boundary conditions are not in doubt, but benchmark cases for models
of intermediate-$\Kn$ flows generally require additional boundary conditions,
usually at solid surfaces, the specification of which is one of the outstanding
problems in hydrodynamic approaches to rarefied gas dynamics. Setting aside the
boundary condition problem, however, we can propose a number of benchmark cases
in categories A and B that any new hydrodynamic model for rarefied flows should
be tested against:
\begin{itemize}
\item the shock structure problem, as outlined in this paper (including
comparisons of $Q_{\rho}$, $\delta_{T\rho}$ and the thickness of $\Ma =\infty$
shocks, in addition to the usual comparison with $L_{\rho}^{-1}$);
\item the nonlinearity of the thermal and momentum Knudsen layers (the region
$O(\lambda)$ close to solid surfaces); \item the `Knudsen paradox' in Poiseuille
flow, i.e.\ the minimum in the mass flow rate at around $\Kn \approx 1$, as well
as bimodality in the temperature profile;
\item drag coefficients and heat transfer in: flow around a sphere, flow around
a cylinder and Couette flow;
\item variation of skin friction on cones in supersonic flow;
\item base pressures on cone-cylinder configurations in supersonic flow as the
Knudsen layer extends far into the wake region;
\item thermophoresis, i.e. the force on particles suspended in a rarefied gas
between two
parallel plates of different temperature;
\item dispersion and absorption of ultrasonic sound waves.
\end{itemize}
This list is neither exclusive nor comprehensive; we are sure that other good
benchmark cases could be added to it. The {\em caveat} is that in most cases
experimental data is extremely sparse and unreliable, and unfortunately much
reliance still needs to be placed on comparison with independent DSMC or other
molecular dynamics simulations as `experimental analogues'.

\section*{Acknowledgements}
We would like to thank Steve Daley of Dstl Farnborough (UK), Howard
Brenner of MIT (USA),  Henry Weller of OpenCFD Ltd.\ (UK), Duncan Lockerby of
Warwick University (UK), and Kokou Dadzie of the University of Strathclyde
(UK) for useful discussions. We also thank the referees of this paper for
their very helpful comments. This work is funded in the UK by the Engineering
and Physical Sciences Research Council under grant GR/T05028/01, and through a
Philip Leverhulme Prize for JMR from the Leverhulme Trust.



\bibliography{hypersonics,general}

\begin{thebibliography}{10}
\expandafter\ifx\csname url\endcsname\relax
  \def\url#1{\texttt{#1}}\fi
\expandafter\ifx\csname urlprefix\endcsname\relax\def\urlprefix{URL }\fi

\bibitem{Brenner:2005a}
H.~Brenner, Kinematics of volume transport, Physica A 349 (2005) 11.

\bibitem{Brenner:2005b}
H.~Brenner, {N}avier-{S}tokes revisited, Physica A 349 (2005) 60.

\bibitem{Schaaf&Chambre:1961}
S.~A. Schaaf, P.~L. Chambr\'e, Flow of rarefied gases, Princeton University
  Press, USA, 1961.

\bibitem{Muckenfuss:1962}
C.~Muckenfuss, Some aspects of shock structure according to the bimodal model,
  Physics of Fluids 5 (1962) 1325.

\bibitem{Bird:1970}
G.~A. Bird, Aspects of the structure of strong shock waves, Physics of Fluids
  13 (1970) 1172.

\bibitem{Barcelo:1971}
B.~T. Barcelo, Electron beam measurements of the shock wave structure: {P}art
  1, {D}etermination of the interaction potential of the noble gases from shock
  wave structure experiments, Ph.D. thesis, California Institute of Technology,
  USA (1971).

\bibitem{Steinhilper:1972}
E.~A. Steinhilper, Electron beam measurements of the shock wave structure: Part
  1, {T}he inference of intermolecular potentials from shock structure
  experiments, Ph.D. thesis, California Institute of Technology, USA (1972).

\bibitem{Sturtevant&Steinhilper:1972}
B.~Sturtevant, E.~A. Steinhilper, Intermolecular potentials from shock
  structure, in: 8th International Symposium on Rarefied Gas Dynamics, 1986, p.
  159.

\bibitem{Alsmeyer:1976}
H.~Alsmeyer, Density profiles in argon and nitrogen shock waves measured by the
  absorption of an electron beam, Journal of Fluid Mechanics 74 (1976) 497.

\bibitem{Narasimha&Das:1986}
R.~Narasimha, P.~Das, The infinitely strong shock, in: V.~Boffi, C.~Cercignani
  (Eds.), 15th International Symposium on Rarefied Gas Dynamics, B. G. Teubner,
  Stuttgart, Germany, 1986, pp. 293--307.

\bibitem{Lumpkin&Chapman:1991}
F.~E. Lumpkin, D.~R. Chapman, Accuracy of the {B}urnett equations for
  hypersonic real gas flows, AIAA Paper 91-0771 (1991).

\bibitem{Zhong&al:1991}
X.~Zhong, R.~W. Mac{C}ormack, D.~R. Chapman, Stabilisation of the {B}urnett
  equations and application to high-altitude hypersonic flows, AIAA Paper
  91-0770 (1991).

\bibitem{Reese:1993}
J.~M. Reese, On the structure of shock waves in monatomic rarefied gases, Ph.D.
  thesis, Oxford University, UK (1993).

\bibitem{Levermore&Morokoff:1998}
C.~D. Levermore, W.~J. Morokoff, The {G}aussian moment closure for gas
  dynamics, SIAM Journal of Applied Mathematics 59 (1998) 72.

\bibitem{Cercignani&Al:1999}
C.~Cercignani, A.~Frezzotti, P.~Grosfils, The structure of an infinitely strong
  shock wave, Physics of Fluids 11 (1999) 2757.

\bibitem{Myong:2001}
R.~S. Myong, A computational method for {E}u's generalized hydrodynamic
  equations of rarefied and microscale gas, Journal of Computational Physics
  168 (2001) 47.

\bibitem{Macrossan&Lilley:2003}
M.~N. Macrossan, C.~R. Lilley, Viscosity of argon at temperatures $>$~2000~{K}
  from measured shock thickness, Physics of Fluids 15 (2003) 3452.

\bibitem{Xu&Tang:2004}
K.~Xu, L.~Tang, Nonequilibrium {B}hatnagar-{G}ross-{K}rook model for nitrogen
  shock structure, Journal of Computational Physics 16 (2004) 3824.

\bibitem{Torrilhon&Struchtrup:2004}
M.~Torrilhon, H.~Struchtrup, Regularized 13--moment equations: shock structure
  calculations and comparison to {B}urnett models, Journal of Fluid Mechanics
  513 (2004) 171.

\bibitem{Struchtrup:2005}
H.~Struchtrup, Macroscopic transport equations for rarefied gas flows,
  Springer, Heidelberg, Germany, 2005.

\bibitem{Maitland&Smith:1972}
G.~C. Maitland, E.~B. Smith, Critical reassessment of viscosities of 11 common
  gases, Journal of Chemical Engineering Data 17 (1972) 150.

\bibitem{Amdur&Mason:1958}
I.~Amdur, E.~A. Mason, Properties of gases at very high temperatures, Physics
  of Fluids 1 (1958) 370.

\bibitem{Aeschliman&Cambel:1970}
D.~P. Aeschliman, A.~B. Cambel, Properties of gases at very high temperatures,
  Physics of Fluids 13 (1970) 2466.

\bibitem{Chapman&Cowling:1970}
S.~Chapman, T.~G. Cowling, The mathematical theory of non-uniform gases, 3rd
  Edition, Cambridge University Press, Cambridge, UK, 1970.

\bibitem{Oettinger:2005}
H.~C. {\"{O}}ttinger, Beyond equilibrium thermodynamics, John Wiley and Sons,
  Hoboken, USA, 2005.

\bibitem{Bardow&Oettinger:2006}
A.~Bardow, H.~C. {\"{O}}ttinger, Consequences of the {B}renner modification to
  the {N}avier-{S}tokes equations for dynamic light scattering, preprint
  submitted to Elsevier Science.

\bibitem{openfoam}
\url{http://www.openfoam.org}.

\bibitem{Issa:1986}
R.~I. Issa, Solution of the implicitly discretised fluid flow equations by
  operator-splitting, Journal of Computational Physics 62 (1986) 40--65.

\bibitem{Rhie&Chow:1982}
C.~M. Rhie, W.~L. Chow, A numerical study of the turbulent flow past an
  isolated airfoil with trailing edge separation, AIAA-82-0998, AIAA/ASME 3rd
  Joint Thermophysics, Fluids, Plasma and Heat Transfer Conference, St. Louis,
  Missouri (1982).

\bibitem{VanLeer:1979}
B.~van Leer, Towards the ultimate conservative difference scheme: {V}, {A}
  second-order sequel to {G}odunov's method, Journal of Computational Physics
  32 (1979) 101--136.

\bibitem{Hirsch:1990}
C.~Hirsch, Numerical computation of internal and external flows, Vol.~2, John
  Wiley and Sons, Chichester, UK, 1990.

\bibitem{Torecki&Walenta:1993}
P.~Torecki, Z.~Walenta, Private communication, {P}olish Academy of Sciences,
  Warsaw, Poland (1993).

\bibitem{Kennard:1938}
E.~H. Kennard, Kinetic theory of gases, McGraw-Hill, New York, USA, 1938.

\bibitem{Struchtrup&Torrilhon:2003}
H.~Struchtrup, M.~Torrilhon, Regularization of {G}rad's 13 moment equations:
  {D}erivation and linear analysis, Physics of Fluids 15 (2003) 2668.

\bibitem{Herwig&Hausner:2003}
H.~Herwig, O.~Hausner, Critical view on ``new results in micro-fluid
  mechanics'': an example, International Journal of Heat and Mass Transfer 46
  (2003) 935.

\end{thebibliography}
\bibliographystyle{elsart-num}

\newpage
\section*{Tables}

\vspace{3em}

\begin{table}[ht]
\centering
\begin{tabular}{|l|c|c|c|c|c|c|} \hline\hline
 & \multicolumn{2}{c|}{{\em Argon}} & \multicolumn{2}{c|}{{\em Helium}} & \multicolumn{2}{c|}{{\em Xenon}} \\ \cline{2-7}
Applicable range & $s$ & $\nu$ & $s$ & $\nu$ & $s$ & $\nu$  \\ \hline
up to Mach 4.3    & 0.68 & 12.0 & 0.71 & 10.7 & 0.77 &  8.5 \\
Mach 4.4--12.5 & 0.76 &  8.8 & 0.83 &  7.2 & 0.72 & 10.3 \\
\hline\hline
\end{tabular}
\caption{Collated experimental values of $s$ and $\nu$ for argon, helium and xenon gases.}
\label{tab:experimentalViscosityTempExponents}
\end{table}

\newpage

\section*{Figures}

\vspace{-1.5em}

\end{document}